\documentclass[submitting]{nst}

\usepackage{subfigure,dcolumn}
\usepackage{epstopdf}
\usepackage{mhchem}
\usepackage{upgreek}
\usepackage{threeparttable}

\begin{document}

\title{Experimental Study of Intruder Components in Light Neutron-rich Nuclei via Single-nucleon Transfer Reaction}
\thanks{Supported by the National Key R\&D program of China (Grant No. 2018YFA0404403), and National Natural Science Foundation of China (Grant Nos. 11775004, U1867214, and 11535004).}

\author{Liu Wei}
\affiliation{School of Physics and State Key Laboratory of Nuclear Physics and Technology, Peking University, Beijing 100871, China}

\author{Lou Jianling}
\email[Corresponding author, ]{jllou@pku.edu.cn}
\affiliation{School of Physics and State Key Laboratory of Nuclear Physics and Technology, Peking University, Beijing 100871, China}
\author{Ye Yanlin}
\affiliation{School of Physics and State Key Laboratory of Nuclear Physics and Technology, Peking University, Beijing 100871, China}

\author{Pang Danyang}
\affiliation{School of Physics, Beijing Key Laboratory of Advanced Nuclear Materials and Physics, Beihang University, Beijing 100191, China}

\begin{abstract}
With the development of radioactive beam facilities, study on the shell evolution in unstable nuclei has become a hot topic. The intruder components, especially $s$-wave intrusion, in the low-lying states of light neutron-rich nuclei near $N$ = 8 are of particular importance for the study of shell evolution. Single-nucleon transfer reaction in inverse kinematics has been a sensitive tool to quantitatively investigate the single-particle-orbital component in the selectively populated states. The spin-parity, the spectroscopic factor (or single-particle strength), as well as the effective single-particle energy can be extracted from this kind of reaction. These observables are often useful to explain the nature of shell evolution, and to constrain, check and test parameters used in nuclear structure models. In this article, we review the experimental studies of the intruder components in neutron-rich He, Li, Be, B, C isotopes by using various single-nucleon transfer reactions. Focus will be laid on the precise determination of the intruder $s$-wave strength in low-lying states.
\end{abstract}


\maketitle

\section{Introduction}\label{sec.I}
Electrons confined by Coulomb potential in atoms possess a well-known shell structure. Shell-like phenomena also appear in finite nuclear system. Almost 70 years ago, Mayer and Jensen succeeded to create a theoretical model to describe the nuclear shell structure \cite{Mayer1, Mayer2}, which was formed by placing the single nucleon (proton or neutron) in a mean field created by all other nucleons. According to the well-established mean field theory, nucleons fill in the single-particle orbitals grouped into shells characterized by the conventional magic numbers, referred to as 2, 8, 20, 50, 82, 126. However, for nuclei far from the $\beta$-stability line, especially those in the light mass region where the concept of a mean field is less robust, the exotic rearrangement of the single-particle configuration often appears and may result in the inversion of some orbitals or the emergence of some intruder components \cite{Chen-PLB}, and further leading to the disappearance of traditional magic numbers and the appearance of new magic numbers \cite{otsuka-2001}. This rearrangement may also result in the shifts of single-particle energies (or orbtials), which further affect essentially all features of the nuclear structure, like deformation \cite{otsuka-2001,otsuka-2005}. The strongly attractive interaction between neutrons and protons has been considered as the main origin of this rearrangement of orbitals in exotic nuclei \cite{otsuka-2001}. Now, the evolution of shell structure in unstable nuclei, including rearrangement of orbitals or intrusion and shifts in single-particle energies, can be partly described in terms of a new mean field model where the monopole effect of the tensor force \cite{otsuka-2005,otsuka-2006,otsuka-2010-1} and three-body forces \cite{otsuka-2010} were implemented.

It has been found that the energy gap between the 1$d_{5/2}$ and 2$s_{1/2}$ shells changes dramatically for light neutron-rich nuclei near $N$ = 8, leading to the appearance of some $s$-wave components in low-lying states. Sometimes, these two orbitals are even inverted, which means the 2$s_{1/2}$ orbital can intrude into 1$d_{5/2}$, and occasionally even intrude into 1$p_{1/2}$  \cite{Ozawa-prl,Kanungo-2013}.
One widely-noted example is the ground state of the one-neutron-halo nucleus $^{11}$Be, which possesses an unusual spin-parity of 1/2$^+$, being dominated ($\sim$71$\%$) by an intruder 2s$_{1/2}$ neutron coupled to an inert core of $^{10}$Be(0$^+$) \cite{Schmitt, Aumann}. Obviously the intrusion of the $s$-wave in the ground state of $^{11}$Be is responsible for the formation of its novel halo structure.  Besides $^{11}$Be,  the intruder $s$-wave components in the low-lying states of $^9$He, $^{10,11}$Li, $^{12,13,14}$Be, $^{13,14}$B, $^{15,16}$C and so on, have also been widely studied by various experiments with different methods in order to understand their exotic structure \cite{Tanihata}. Transfer reaction is one of the most commonly-used experimental methods to study such kinds of intruder components in exotic nuclei.

Transfer reaction, especially the single-nucleon transfer reaction, is a sensitive experimental tool to populate a certain interesting state in nuclei with a selective manner. These populated states can be described by an original or a residual nucleus as a core with the transferred nucleon moving around it in a certain orbital. The spin-parity of the selectively populated state can be assigned because the oscillation behaviour of differential cross sections (DCSs) depends on the transferred angular momentum $l$ \cite{Alexandre,Wimmer}. The contribution of the transferred nucleon to each populated state is usually described by the spectroscopic factor ($SF$), which is often extracted by comparing the experimental DCSs to the theoretical ones. It has been disputed for a long time if the $SF$ is a good experimental observable or not. Can we connect the $SF$ with the nuclear structure directly? In order to answer these questions, many experiments have been performed. Conclusions have been reached that it is the relative or the normalized $SF$ instead of the directly extracted one (reaction model dependent), that is related to the occupancy or vacancy number of a certain orbital \cite{Schmitt,Kay-2013}. If both the $SF$ and the excitation energy for each populated state with the same orbital $j$ are known, the effective single-particle energy for the orbital $j$ can also be determined \cite{Alexandre,Wimmer}. Therefore, transfer reaction is a quantitative tool to probe the intensity of the single-particle-orbital component (or occupancy/vacancy number), the spin-parity of the populated state, and the effective single-particle energy for the nuclei far away from the $\beta$-stability line.

 Unlike stable nuclei,  the half-lives of radioactive beam are relatively short, thus single-nucleon transfer reactions have to be performed in inverse kinematics. The missing mass method, in which only the energies and angles of the recoil light particles are measured,  is a commonly used method for transfer reactions in inverse kinematics. This experimental method has some advantages over the normal kinematic measurements.  However, at the same time, it also introduces a number of experimental challenges, such as the precise detection of the charged-particles with very low energy. Therefore, some new experimental techniques are developed to overcome these disadvantages. Many new detection arrays are constructed in various laboratories. In this paper, the basic modules, advantages and disadvantages of the typical experimental setups are introduced in details. Along with these setups, the studies of single-particle-orbital intruder component in light neutron-rich nuclei are reviewed.

This paper is organised as follows. In section. \ref{transfer reaction}, the basic concepts of transfer reaction are outlined.  In section. \ref{experiment}, the advantages and disadvantages in normal and inverse kinematics are analyzed, and the missing mass method is introduced.  In section. \ref{detector}, typical experimental setups for single-nucleon transfer reaction used worldwide are summarized. In section. \ref{experimentalresults}, the studies of exotic nuclear structure of the neutron-rich He, Li, Be, B, C isotopes using single-nucleon transfer reactions are reviewed, and a brief summary is given in the last section.

\section{What can we learn from the transfer reaction?}\label{transfer reaction}
\subsection{What is a transfer reaction?}
\begin{figure}[htb]
\includegraphics
  [width=0.9\hsize]
  {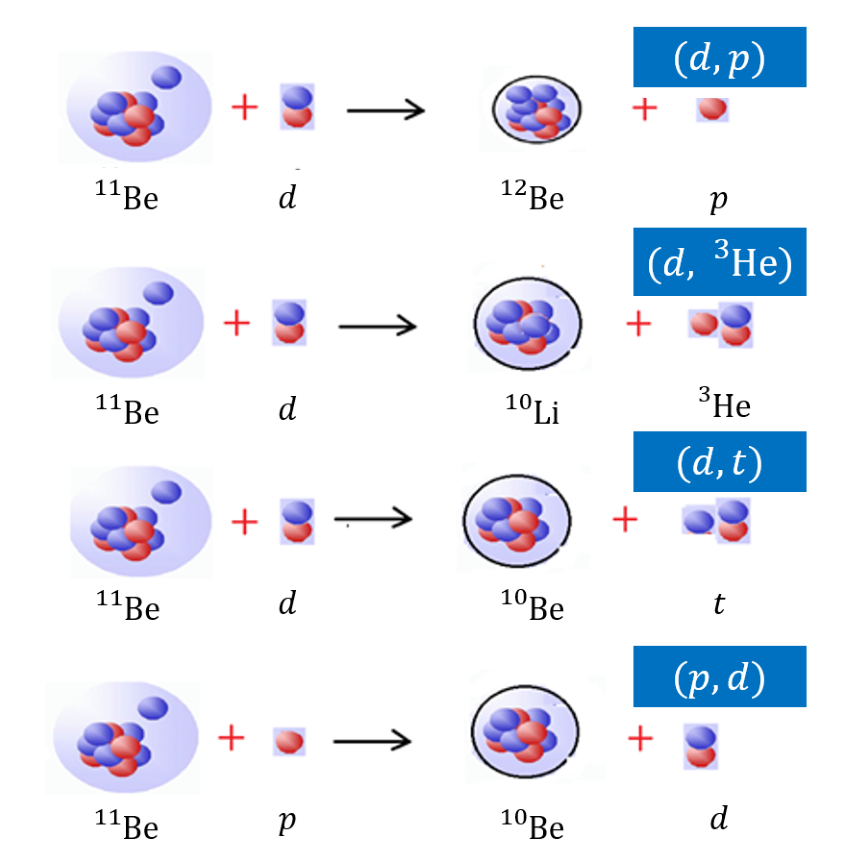}
\caption{Single-nucleon transfer reactions induced by $^{11}$Be impinging on the proton or the deuteron target.}
\label{single-nucleon-transfer}
\end{figure}

Generally, a transfer reaction can be written as
\begin{equation}
{\rm {A+a \rightarrow B+b \qquad or \qquad A(a,b)B,}}
\end{equation}
where A and a stand for the target and the projectile nuclei, while B and b represent the residual and the outgoing particles, respectively. It means that when the projectile a impinges on the target A, a nucleon, proton or neutron, or a cluster (such as $^4$He) was transferred to form a new final system comprised of b and B. If the nucleon or cluster is removed from the projectile a to the target A, it
is called a stripping reaction, when it is added to the projectile a, the reaction is called a pickup reaction. If only one nucleon was transferred, we call it single-nucleon transfer reaction.  Fig. \ref{single-nucleon-transfer} shows different kinds of single-nucleon transfer reactions induced by $^{11}$Be colliding on the proton or deuteron target. For a beam of deuteron impinging on a $^{11}$Be target, the one-neutron and one-proton transfer reactions were written as $^{11}$Be($d$, $p$)$^{12}$Be (or $^{11}$Be($d$, $t$)$^{10}$Be) and $^{11}$Be($d$, $^3$He)$^{10}$Li, respectively. In the case of a $^{11}$Be incidence, the experiment has to be performed in inverse kinematics, the corresponding expressions are changed into  $d$( $^{11}$Be, $p$)$^{12}$Be (or $d$($^{11}$Be, $t$)$^{10}$Be) and $d$($^{11}$Be, $^3$He)$^{10}$Li. In both cases, the recoil charged-particle proton (or triton) and $^3$He are measured. In the case of inverse kinematics, both the reactions of $d$($^{11}$Be, $t$)$^{10}$Be and $d$($^{11}$Be, $^3$He)$^{10}$Li are typical stripping or nucleon-removing reactions, while the $d$($^{11}$Be, $p$)$^{12}$Be reaction belongs to pickup or nucleon-adding reaction. The single-particle knock-out reaction is another typical kind of nucleon-removing reaction, but will not be reviewed in this paper.

\subsection{Angular distributions and spin-parity}
\begin{figure}[htb]
\includegraphics
  [width=0.9\hsize]
  {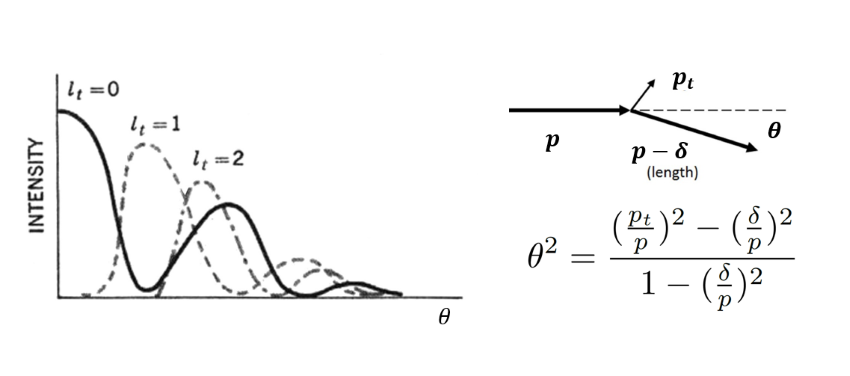}
\caption{In the left-hand side, the DCSs as a function of the
scattering angle $\theta$ in lab frame for different transferred
orbital angular momentum $l$. In the right-hand side, a vector diagram to explain
the relationship between $\theta$ and the
transferred momentum $l$. See more details in text. This figure is from Ref. \cite{Catford}.}
\label{spinparity}
\end{figure}

The DCS of A(a, b)B to a given final state in B, as a function of the scattering angle, shows an oscillation behaviour. The oscillated structure (positions of maxima and minima) depends on the transferred angular momentum $l$.
This behavior can be understood by a simple momentum diagram shown in Fig. \ref{spinparity}. If we assume that the incident projectile has a momentum $\vec{p}$ and the momentum transferred to the target nucleus is $\vec{p}_t$, then the
beam particle will have only a small reduction in the magnitude of its momentum for a small scattering angle $\theta$, as seen in the vector diagram (Fig. \ref{spinparity}) built according to the momentum conservation. From the cosine rule, we have
\begin{equation}
{\cos}\theta=\frac{p^2+(p-\delta)^2-p_t^2}{2p(p-\delta)}.
\end{equation}
If we make use of the expansion to the second order, referred to as,
\begin{equation}
  \rm{\cos}\theta
  \approx
  1-\frac{\theta^2}{2} ,
\end{equation}
we have,
\begin{equation}
\theta^2=\frac{(\frac{p_t}{p})^2-(\frac{\delta}{p})^2}{1-(\frac{\delta}{p})^2}.
\end{equation}
The reduction $\delta$ in the length of the
vector $p$ is small in comparison to the length of the actual transferred momentum $p_t$. Hence, we can drop the term of $\delta/p$ \cite{Catford}. Then, the expression of $\theta^ 2$ is simplified to
\begin{equation}
\theta^ 2\approx (\frac{p_t}{p})^2.
\end{equation}
In the classical picture of transfer reaction, the nucleon is usually transferred at the surface of the target nucleus, then the angular momentum $\vec{L}$ is given by \cite{Catford}
\begin{equation}
\vec{L} = \vec{p}_t\times \vec{R},
\end{equation}
where $\vec{R}$ is the vector radius of target nucleus. At the same time, from quantum mechanics \cite{Alexandre}, we have
\begin{equation}
{L}^2\mid\phi> = l(l+1)\hbar^2 \mid \phi>.
\end{equation}
Therefore, it is easy to deduce
\begin{equation}
\theta_0 \approx \frac{p_t}{p} = \frac{\sqrt{l(l+1)}\hbar}{pR},
\end{equation}
where $\theta_0$ is the angle of first maxima of the cross section. This indicates that $\theta_0$ is different for different transferred momentum $l$, and $\theta_0$ increases with $l$.  This trend can be seen in Fig. \ref{spinparity}.  The transferred orbital angular momentum $l$, and therefore the parity of the populated states, can be assigned in conjunction with other experimental information or shell model predictions. Furthermore, transfer reactions
can be used to assign the total angular momentum $j$ of the selectively populated states as well \cite{Wimmer}.

\begin{figure}[htb]
\includegraphics
  [width=0.98\hsize]
  {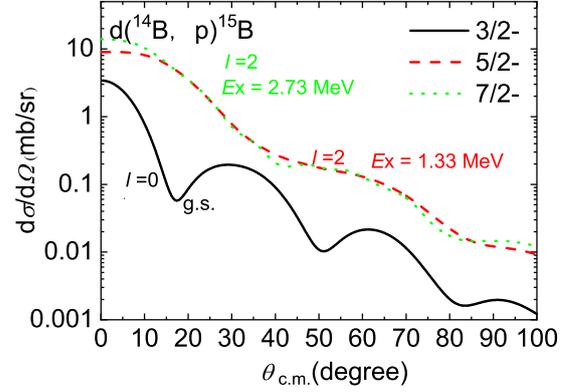}
\caption{Angular distributions for the $^{14}$B($d$, $p$) transfer reaction to the 3/2$^-$ ground state (solid curve), the 5/2$^-$ (dashed curve) and 7/2$^-$ (dotted curve) excited states in $^{15}$B with a radioactive beam of $^{14}$B at 25 MeV/nucleon. The curves are calculated by using the code FRESCO \cite{Fresco} and systematical optical potentials.}
\label{angular-distribution}
\end{figure}

 Take the reaction of $^{14}$B($d$, $p$) for example, Fig. \ref{angular-distribution} depicts angular distributions for this reaction to the 3/2$^-$ ground state (solid curve), the 5/2$^-$ (dashed curve) and 7/2$^-$ (dotted curve) excited states in $^{15}$B. The DCSs in Fig. \ref{angular-distribution} were calculated from the code FRESCO \cite{Fresco} with the global optical potentials obtained from W. W. Daehnick \cite{Daehnick} and A. J. Koning $\&$ J. P. Delaroche (KD02) \cite{KD02} for the entrance and exit channel, respectively. It is obvious that the oscillation behaviour of angular distributions largely depend on the transferred angular momentum $l$.

\subsection{Spectroscopic factor}
 Single-nucleon transfer reaction is a powerful experimental tool to study the structure of a certain selectively populated states. The contribution of the transferred nucleon to each populated state is usually described by the $SF$, which is often determined by comparing the DCSs deduced from experiments with those from theoretical calculations. The functional expression is as follows.

\begin{equation}\label{sfequation}
(\frac{d \sigma}{d \Omega})_{\textrm{exp}} = C^2 SF_{\textrm{exp}}(\frac{d \sigma}{d \Omega})_{\textrm{theo}},
\end{equation}
where $C^2$ is the isospin Clebsch–Gordan coefficient, which is sometimes omitted. It amounts to 1.0 for the ($d$, $p$) transfer reaction \cite{Wimmer}. $(\frac{d \sigma}{d \Omega})_{\textrm{exp}}$ and $(\frac{d \sigma}{d \Omega})_{\textrm{theo}}$ are the DCSs, which were extracted from experiments and various reaction models, respectively. It should be noted that the expressions of ($\frac{d \sigma}{d \Omega})_{\textrm{theo}}$ from different reaction codes are often different. For example, from the code FRESCO \cite{Fresco} and DWUCK \cite{DWUCK}, they are
\begin{equation}\label{frescoanddwuck}
 (\frac{d \sigma}{d \Omega})_{\textrm{theo}}
= (\frac{d \sigma}{d \Omega})_{\textrm{FRESCO}}
=\frac{2J_f+1}{2J_i+1} (\frac{d \sigma}{d \Omega})_{\textrm{DWUCK}},
\end{equation}
where $J_i$ and $J_f$ are the spins of initial and final nuclei.

The Distorted Wave Born Approximation (DWBA) is the most commonly-used approximation theory for the calculations of the transfer reaction DCSs ($(\frac{d \sigma}{d \Omega})_{\textrm{theo}}$). For a given reaction, the theoretical DCS is given by
\begin{equation}
(\frac{d \sigma_{\beta\alpha}}{d \Omega})_{\textrm{theo}}
=\frac{\mu_\alpha\mu_\beta}{(2\pi\hbar^2)^2}
\frac{\kappa_{\beta}}{\kappa_{\alpha}}
|T_{\beta\alpha}(\kappa_\beta, \kappa_\alpha)|^2
\end{equation}
where $\mu_\alpha$ ($\kappa_\alpha$) and $\mu_\beta$ ($\kappa_\beta$) are the reduced masses (wave numbers) in the entrance and exit channel, respectively. $T_{\beta\alpha}$ is the transition amplitude.
It takes into account the
distortion of the incoming and outgoing waves caused by the nuclear potential $U$ between the projectile and the target
in the entrance (a + A, $\alpha$) and exit partitions (b + B, $\beta$). Assuming that the transfer reaction occurs in one step (first-order DWA or DWBA), the transition amplitude $T_{\beta\alpha}$ can then be written as \cite{Alexandre}
\begin{equation}
T_{\beta\alpha} =\int
\chi^{(-)}
{(k_\beta, r)}
<\Phi_{\beta}|\Delta U|\Phi_{\alpha}>
\chi^{(+)}
(k_{\alpha}, r)
\textrm{dr},
\end{equation}
where $r$ is the relative distance between the projectile and the target (${r_{\alpha(\beta)}}$ in the entrance (exit) channel),
 $\chi^{(+)}{(k_{\alpha},r)}$
is the “distorted” wave composed of an incoming plane wave in the state $\alpha$ and outgoing scattered waves. Similarly,
 $\chi^{(-)}{(k_\beta,r)}$
is the outgoing distorted wave in the channel $\beta$. The waves $\chi^{(-)}
{(k_\beta, r)}$ and $\chi^{(+)}
(k_{\alpha}, r)$ are obtained by solving the Schrodinger
equation with an assumed potential for outgoing and incoming channels, respectively. The potentials are usually extracted from the elastic scattering DCSs using the optical model (OM). The extracted potentials are called optical model potentials (OPs).

In the case of reaction on unstable nuclei, the elastic scattering
data do not always exist, leading to larger uncertainties
in the DWBA calculations. Moreover, as the deuteron is
relatively loosely bound (only 2.22 MeV), it is easy to break
up in the presence of target nucleus. This breakup channel can couple to the transfer channel, affecting
the $SF_{{\rm exp}}$ extracted in a nontrivial manner.
To account for this mechanism, Johnson and Soper \cite{Johnson-Soper}
devised the adiabatic wave approximation (ADWA), which
uses nucleonic potentials and explicitly includes deuteron
breakup. An extension of this method to include finite
range effects (FR) was introduced by Johnson
and Tandy \cite{Johnson-Tandy}. As stated in Ref. \cite{Schmitt}, for the ($d$, $p$) reactions, the $SF$s extracted using the adiabatic model (FR-ADWA) are “stable” across measurements at four energies and are insensitive to the applied OPs, demonstrating its advantages in comparison with the extraction using the normal DWBA approach. This is why the FR-ADWA model was adopted for most analysis of transfer reaction now. It is worth noting that those four measurements were performed under almost identical experimental conditions and used the same set of OP parameters \cite{Schmitt}.

 The experimentally extracted $SF$ is sensitive to the choice of the applied OP and also to some practical experimental conditions \cite{Schmitt,Kay-2013}. For a long time, it has been disputed whether it is a good experimental observable.
Conceptually the $SF$ is used to describe the occupancy of a valence nucleon at a single-particle orbit in a mean field created by other nucleons. For instance, at an orbit with spin $j$, the total degeneracy number within the independent-particle model (IPM) should be 2$j$ + 1. In the case of configuration mixing based on the shell model approach, this number may be split into several states which are composed of the same $j$-wave with a certain intensity. Shell model calculations with appropriate effective interactions and model space could, in principle, predict the $SF$ of a certain wave (single-particle orbit) in an energy-eigen-state. But the experimentally observed $SF$s are often smaller than the shell-model predictions, an effect being exhibited by a reduction or quenching factor. This quenching phenomenon was firmly established from (e, e’$p$) knockout reactions \cite{eelastic1,eelastic2}. Using nuclear reactions, such as single-particle knockout or transfer reactions, this quenching effect is also generally confirmed. Using these reactions, the $SF_\textrm{exp}$ can be extracted by comparing the experimentally measured cross section with the calculated one assuming a pure single-particle state \cite{Pain,Kanungo} (Eq.(\ref{sfequation})). Since the individual $SF_{{\rm exp}}$ might be sensitive to the choice of OPs and to the practical experimental conditions (see above), the sum rule method was developed to define the relative $SF$ and the general quenching factor \cite{sumrule}. The quenching factor in the nucleon-transfer reaction is defined as  \cite{Kay-2013}

 \begin{equation}
F_q = \frac{1}{2j+1}[\sum(\frac{\sigma_{\textrm{exp}}}{\sigma_{\textrm{theo}}})^\textrm{rem}_j+\sum(\frac{\sigma_{\textrm{exp}}}{\sigma_{\textrm{theo}}})^\textrm{add}_j],
\end{equation}
where the sum of the adding and removing relative cross sections for a given $l$, $j$ represents the total degeneracy (sum-rule) of that orbit.

For the radioactive beam, it is hard to measure the nucleon-removing and nucleon-adding reactions at the same experiment due to the limited beam intensity. If only the nucleon-adding or nucleon-removing data were available for a given nucleus, the function for the quenching factor requires that the total
strength adds up to the number of vacancies in the closed
shell, or the number of particles outside it \cite{Kay-2013}, and the quenching factor is modified to
 \begin{equation}
F_q = \frac{1}{2j+1}[\sum(\frac{\sigma_{\textrm{exp}}}{\sigma_{\textrm{theo}}})_j].
\end{equation}

For instance, Ref. \cite{Kay-2013} reported a consistent quenching factor of about 0.55 for a large number of nuclei, with a root-of-mean-square spread of 0.10.
Once the sum rule was established, the individual $SF_{\textrm{exp}}$ can be normalized through the sum rule to give the intensity (percentage) of the wave component. This normalized $SF$ is much less sensitive to the employed OPs etc. and can be reasonably used to compare with the theoretical predictions.
In other words, even for the same reaction, it is hard to directly compare the experimental $SF$s extracted from different measurements and analyzed using various sets of OP options without any normalization procedure \cite{Kay-2013}. Only the intensities (percentages) or the normalized $SF$ (also called relative $SF$ in some references), not the directly extracted experimental $SF$s, of the wave components, can be compared directly with each other.

It is worth noting that the $SF$ normalization procedure does not change the ratio between the $SF_{\textrm{exp}}$ of different populated states in final nucleus with the same spin-parity. Hence the ratio, which is equivalent to the $SF$ normalization, is often used in experiments \cite{Chen-PLB,16C-Wuosmaa}.

\subsection{Effective single-particle energy}
The IPM assumes that nucleons lie on single-particle energy orbits with no correlation among them. Therefore, the single-particle energy of a certain orbit/state, which is equal to the energy needed for one nucleon to be excited from the ground state to this orbit/state, can be simply measured. The shell model is based on a mean-field theory, in which the correlations between nucleons (protons and neutrons) are taken into consideration \cite{Alexandre}. In this case, the single-particle (uncorrelated) energies are not direct observables since real nuclei are correlated systems by nature, but they can be obtained from experimental data using the normalized $SF$ and excitation energy of each populated state according to the sum rule \cite{sumrule}.

Based on the Macfarlane-French sum rule \cite{sumrule},
for the nucleon-removing reaction of a given nucleus A, the number of nucleons populated in a shell $j$ (occupancy number) of A, $ G^-(j)$, is
\begin{equation}\label{G-}
  G^-(j)=\sum_k (SF)_k.
\end{equation}
For the neutron-adding reaction of a nucleus A, the number of holes in a shell $j$ (vacancy number) of A, $ G^+(j)$, is
\begin{equation}\label{G+}
  G^+(j)=\sum_k \frac{(2J_f+1)_k}{2J_i+1} (SF)_k,
\end{equation}
where $J_i$ and $J_f$ is the spin of initial and final state ($k$), respectively. It should be noted that Eq. (\ref{G-}) and Eq. (\ref{G+}) are simple ones for the spin zero target, such as the deuteron target. This sum rule has been tested in the neutron transfer reactions by J. P. Schiffer $et$ $al$ \cite{eelastic2}.
Neutron-adding, neutron-removal, and proton-adding transfer reactions were measured
on the four stable even Ni isotopes, with particular attention to the cross section determinations. They found that valence-orbit occupancies extracted from the sum rule, are consistent with the changing number of valence neutrons, as are the vacancies
for protons, both at the level of $<5\%$ \cite{eelastic2}. This sum rule has also been used in the reactions of
$d$($^{13}$B, $p$) \cite{Bedoor},
 $d$($^{19}$O, $p$) \cite{Hoffman-19O},
 as well as $d$($^{12}$B, $^{3}$He) \cite{Chen-12B} and so on.

If excitation energies and spectroscopic strengths, referred to as the normalized or the relative $SF$s, for all
the relevant states with the same transferred angular momentum $l$ are available, the effective single-particle energies (ESPE) are given by the centroid \cite{Alexandre}:
\begin{equation}\label{ESPE}
  \varepsilon(j) = \frac{\sum_{k}G^+(E^+_k-E_0) + G^-(E_0-E^+_k) }{G^+ + G^-},
\end{equation}
where the sum is over all final excited states $k$, $ \varepsilon(j) $ is the ESPE, and $E^{\pm}_k$  is the excitation energy for the state $k$ in nucleus $A \pm 1$. $E_0$ is the ground state energy of the nucleus A.
For the adding-nucleon reaction, such as ($d$, $p$), we have
\begin{equation}\label{ESPE1}
  \varepsilon (j) = \frac{\sum_{k} \frac{(2J_f+1)_k }{(2J_i+1)}(SF)_k E_k}{G^+(j)},
\end{equation}
where $SF_k$ is the relative or normalized $SF$ for the state $k$. It can be simplified as
\begin{equation}\label{ESPE2}
  \varepsilon (j) = \frac{\sum_{k} (2J_f+1)_k (SF)_k  E_k}{\sum_{k}(2J_f+1)_k (SF)_k}.
\end{equation}
This formulation is from Ref. \cite{Bedoor} and is equivalent to
that given in Ref. \cite{Baranger}.

\begin{figure}[htb]
\includegraphics
  [width=0.9\hsize]
  {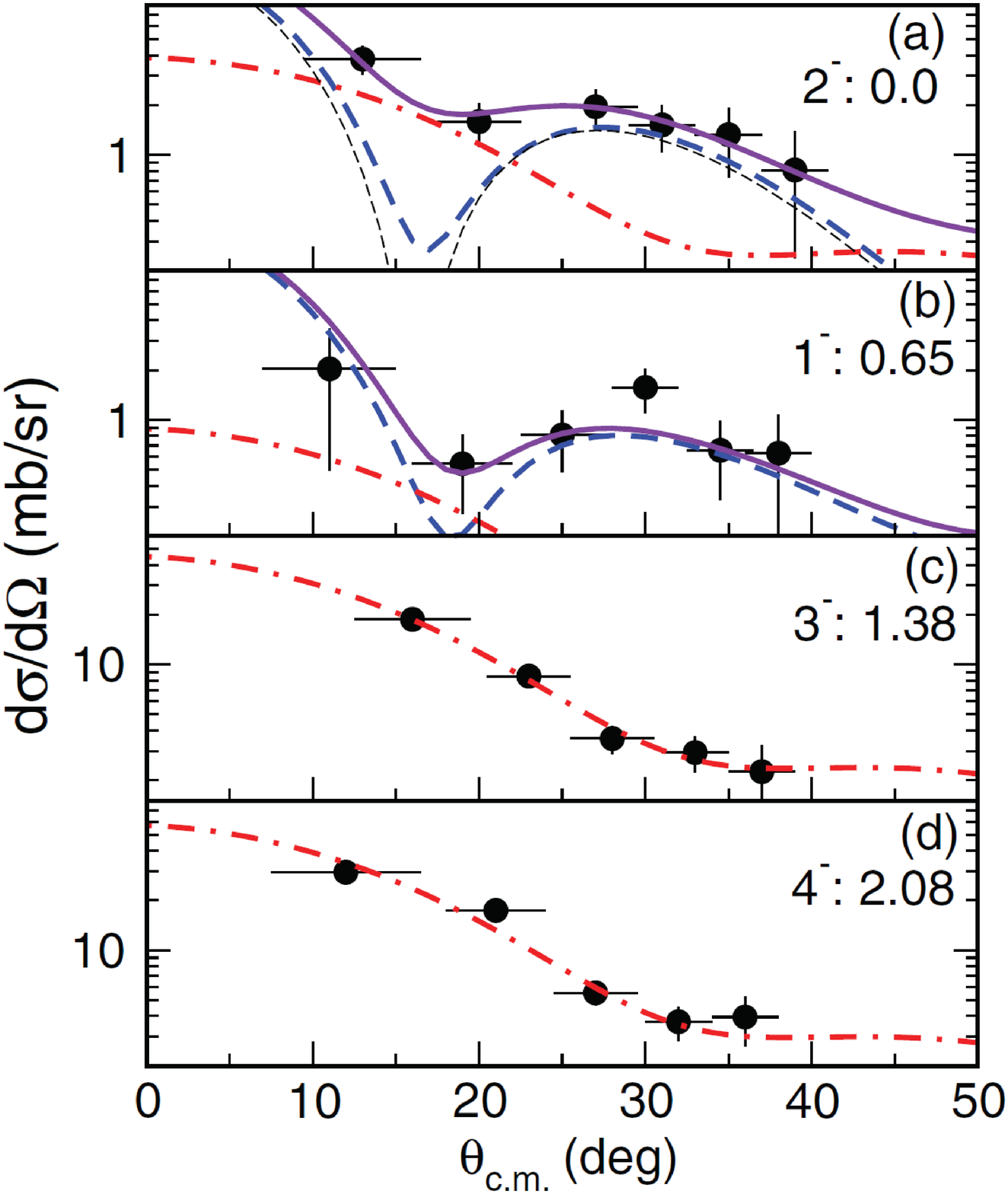}
\caption{The DCSs for
the $^{13}$B($d$, $p$) reaction to different excited states in $^{14}$B. The horizontal bars represent the angular
range for each data point, while the vertical bars stand for the statistical error. The thick-dashed, dot-dashed, and solid
curves are corresponding to the DWBA calculations with $l$ = 0, 2, and 0 + 2, respectively. The thin-dashed
curve in (a) shows the $l$ = 0 result for the 2$_1^-$ state before
averaging over the scattering angle. This figure is from Ref. \cite{Bedoor}.}
\label{13bdpangular}
\end{figure}

Take for example the reaction of $d$($^{13}$B, $p$) to the low-lying states in $^{14}$B, let us show how to calculate the ESPE using Eq. (\ref{ESPE2}). The populated low-lying states in $^{14}$B were constructed by
the coupling of one $sd-$shell neutron to the
 $3/2^-$ ground
state of $^{13}$B. This coupling leads to a $(1,2)^-$ doublet for
$1\pi(1p_{3/2})$-
$1\nu(2s_{1/2})$,
and (1,2,3,4)$^-$ and (0,1,2,3)$^-$ multiplets
when the transferred neutron populates the 1$d_{5/2}$ and 1$d_{3/2}$ orbitals, respectively. Configuration
mixing between states with the same spin and parity in
$^{14}$B is possible, especially for the neutron 2$s_{1/2}$ and 1$d_{5/2}$
orbitals due to the smaller energy gap between these two orbitals.  Fig. \ref{13bdpangular} shows the angular distributions for the $d$($^{13}$B, $p$) reaction to four low-lying states in $^{14}$B. Only the 2$_1^-$, 1$_1^-$, 3$_1^-$, and 4$_1^-$ states were populated, therefore only the $SF$s for these four states were extracted from this experiment. In Fig. \ref{13bdpangular}(a) and Fig. \ref{13bdpangular}(b), it was found that it is hard to fit the 2$_1^-$ and 1$_1^-$ DCSs only using the $s$-wave ($l$ = 0) components (blue thick-dashed curves). However, when a little $d$-wave component was taken into consideration, the angular distributions can be fitted better, indicating that these two states are made up of $s$- and $d$-wave mixture.  If we ignored the effect of 1$d_{3/2}$ orbital, the spectroscopic strengths or the normalized $SF$s for the unobserved 2$_2^-$ and 1$_2^-$ states can be deduced with an assumption that the pairs of $2^-$ and $1^-$ levels are formed by
orthogonal combinations of 2$s_{1/2}$ and 1$d_{5/2}$ configurations. According to the orthogonal rule, the wave functions for these two states are written as
\begin{equation}\label{orthogonal}
\begin{split}
|J_1^-\rangle= \alpha_J\nu (2s_{1/2}) + \beta_J \nu (1d_{5/2}),\\
|J_2^-\rangle= - \beta_J\nu (2s_{1/2}) + \alpha_J \nu (1d_{5/2}),
\end{split}
\end{equation}
where $J$ = 1 and 2, $\alpha_J\times\alpha_J$ = $SF(l=0)$ and $\beta_J$$\times \beta_J$ = $SF(l=2)$ for the $2_1^-$ (or $1_1^-$) state, as well as $\beta_J\times\beta_J$ = $SF(l=0)$ and $\alpha_J$$\times \alpha_J$ = $SF(l=2)$ for the $2_2^-$ (or $1_2^-$) state.
According to the results of experimentally observed 2$_1^-$ and 1$_1^-$ states, the $s$-wave ($d$-wave) $SF$s are easily determined to be 0.17(5)(4) (0.71(5)(20)) and $\leq$0.06 (0.94(20)(20)) for the unobserved 2$_2^-$ and 1$_2^-$ states, respectively. This simple orthogonal method has been widely used in experiments, such as $d$($^{15}$C, $p$) \cite{16C-Wuosmaa}, $d$($^{11}$Be, $p$) \cite{Chen-PLB}.

The excitation energies ($E_x$) and $SF$s for each populated state are listed in Tab. \ref{13bdp} \cite{Bedoor}. Using Eq. (\ref{G+}), the holes are determined to be 1.9 $\pm$ 0.2 and 5.9 $\pm$ 0.3 for $2s_{1/2}$ and $1d_{5/2}$ orbitals, respectively. These two values are very close to numbers of 2.0 and 6.0 predicted by IPM. With Eq. (\ref{ESPE}), the ESPE for the $2s_{1/2}$ and $1d_{5/2}$ orbitals are deduced to be about 0.5 $\pm$ 0.1 and 2.0 $\pm$ 0.4 MeV, respectively. These results demonstrate that the $s$-orbital is lower than $d$-orbital in $^{14}$B, which is different from the orbital arrangement in conventional shell model but is similar to other $N$ = 9 isotones $^{13}$Be and $^{15}$C.

\begin{table}[!htb]
\caption{The excitation energies and normalized $SF$s for the low-lying states in $^{14}$B \cite{Bedoor}. The $SF$s are normalized to $SF(3^-)$ = 1.0 and the uncertainties are (fit)(theory). The $SF$s in middle brackets are calculated from the experimentally observed ones with an assumption that the pairs of 2$^-$ and 1$^-$ levels are formed by orthogonal combinations of 2$s_{1/2}$ and 1$d_{5/2}$ configurations, see more details in text.}
\label{13bdp}
\begin{tabular*}{8cm} {@{\extracolsep{\fill} } cccc}
\toprule
spin-parity& $E_x$(MeV)&$SF$($l=0$)&$SF$($l=2$)\\\hline
2$_1^-$ & 0 &0.71(5)(20)&0.17(5)(14)\\
1$_1^-$ & 0.654 &0.94(20)(20)&$\leq$0.06\\
3$_1^-$ & 1.38 &&$\equiv$ 1.0\\
2$_2^-$ & 1.86 &[0.17(5)(4)]&[0.71(5)(20))]\\
4$_1^-$ & 2.08 &&1.0\\
(1$_2^-$) & 4.5 &$\leq$0.06&[0.94(20)(20)]\\\hline
Eq.(\ref{G+})&&1.9 $\pm$ 0.2&5.9 $\pm$ 0.3\\
\bottomrule
\end{tabular*}
\end{table}

\section{Experimental methods} \label{experiment}
\subsection{Normal kinematics}
\begin{figure}[htb]
\includegraphics
  [width=0.98\hsize]
  {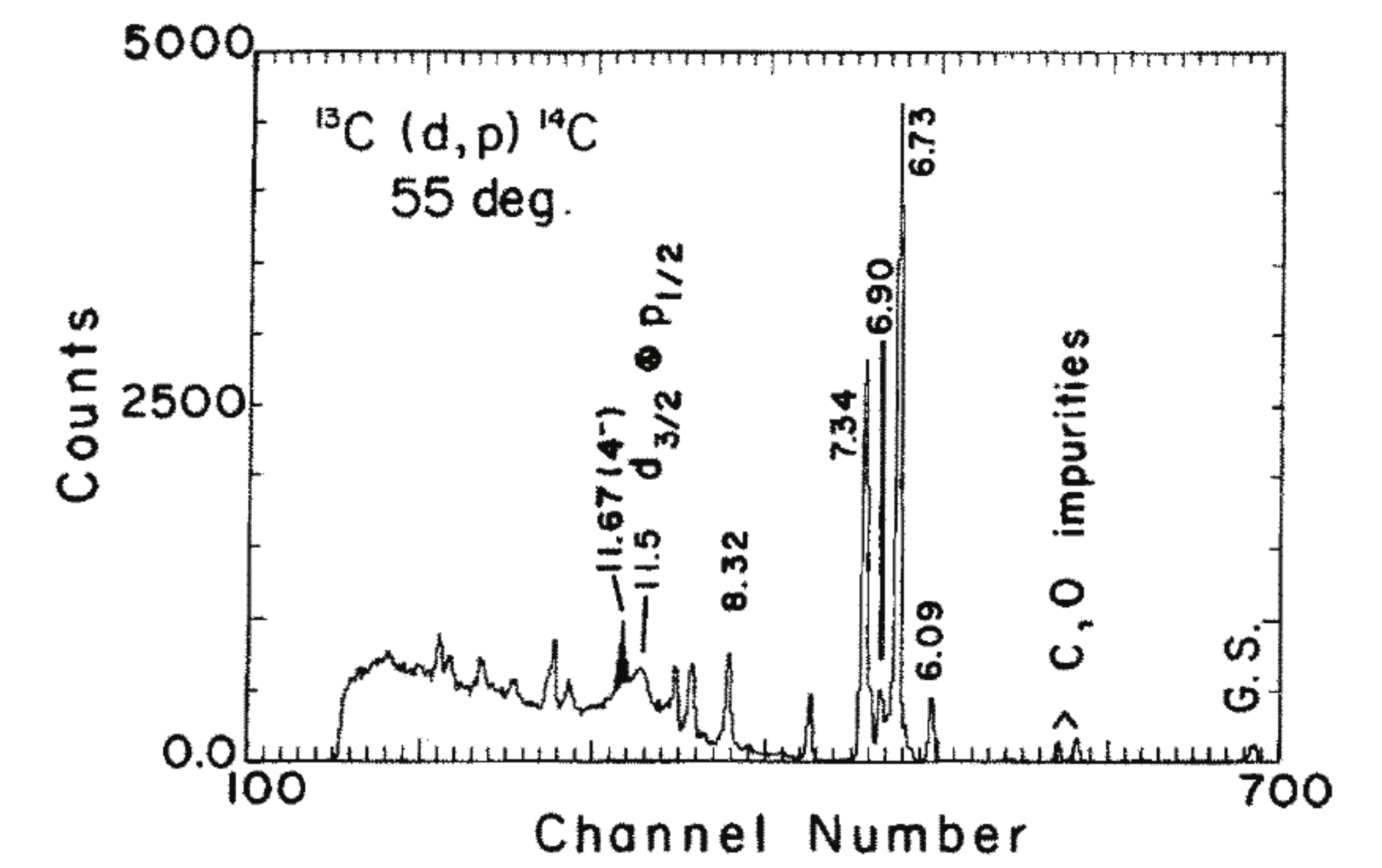}
\caption{The $Q-$value spectrum for $^{13}$C($d$, $p$)$^{14}$C with a deuteron beam at 17.7 MeV in normal kinematics \cite{Peterson-13cdp}.}
\label{13cdpnormalkinematics}
\end{figure}

When using a beam of deuteron and a stable target, one of the best ways to measure ($d$, $p$), ($d$, $t$) and ($d$, $^3$He) reactions is to use a high resolution magnetic spectrometer to record the recoil charged particles $p$, $t$ and $^{3}$He emitted from these reactions. High precision and low background are two typical advantages of normal kinematics.
The typical energy resolution of the excited states in final nucleus is about several tens of keV. For example, the $Q-$value spectrum for the $^{13}$C($d$, $p$)$^{14}$C reaction with a deuteron beam at 17.7 MeV in normal kinematics is shown in Fig. \ref{13cdpnormalkinematics} \cite{Peterson-13cdp}. It was found that three closely spaced excited states at $E_x$ = 6.73, 6.90, and 7.34 MeV in $^{14}$C are clearly identified, and the average resolution is about 60 keV \cite{Peterson-13cdp}.

\subsection{Inverse kinematics}
For the radioactive beams, whose half-lives are relatively short and separation energies are relatively low, it is hard to use the normal kinematics because it is nearly impossible to use them as targets. In this case, the inverse kinematics is usually applied, in which the deuteron or the proton is used as target while the radioactive beam is projectile.

\begin{figure}[htb]
\includegraphics
  [width=0.9\hsize]
  {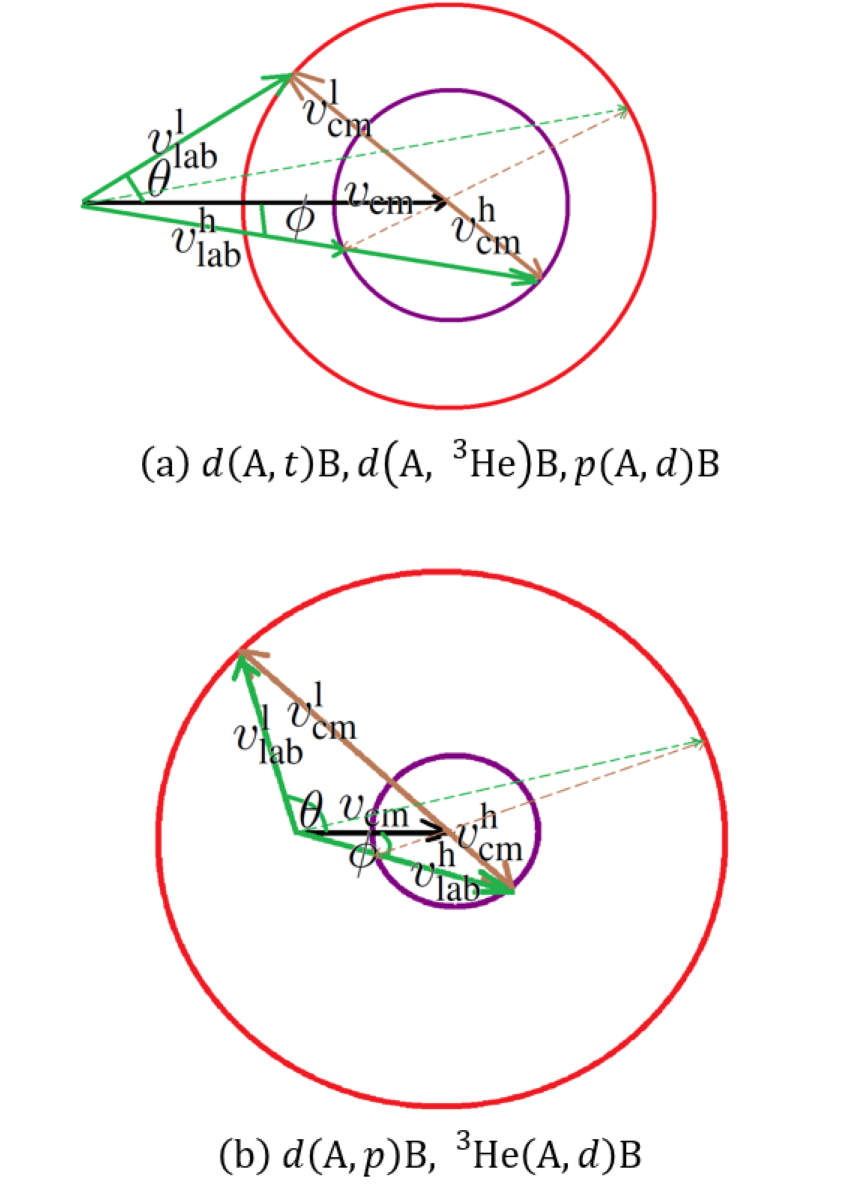}
\caption{In the case of inverse kinematics, vector diagrams for the reactions of (a) $d$(A, $t$)B, $d$(A, $^3$He)B, $p$(A, $d$)B and (b)$d$(A, $p$)B and $^3$He(A, $d$)B. The large and small circles represent the possible vector scope for light and heavy particles in CM frame, respectively. The thick solid and thin dashed lines stand for the low- and high-energy branch of light product, respectively.}
\label{kinematicsininverse}
\end{figure}

The vector diagram for the reaction A(a, b)B in inverse kinematics is shown in Fig. \ref{kinematicsininverse}. The centre of
mass (CM) vector $v_\textrm{cm}$ has the same direction as the projectile, and its length is
\begin{equation} v_\textrm{cm}=\frac{m_\textrm{a}}{m_\textrm{a}+m_\textrm{A}}\times v^\textrm{lab}_\textrm{a},
\end{equation}
where, $m_\textrm{a}$ ($v^\textrm{lab}_\textrm{a}$) and $m_\textrm{A}$ are the mass (velocity) of projectile and target, respectively.
In the case of (a) $d$(A, $t$)B, $d$(A, $^3$He)B, or $p$(A, $d$)B, the vector diagram is given in Fig. \ref{kinematicsininverse}(a). After reaction, the heavy particle is going
to the forward angles with little change in velocity and direction. According to momentum and energy conservation, it is easy to obtain a rough estimate of the vector
length of the light (heavy) particle in the CM frame, labelled as $v^\textrm{l}_\textrm{cm}$ ($v^\textrm{h}_\textrm{cm}$) in the Fig. \ref{kinematicsininverse}(a). Take the reaction of $p$(A, $d$)B for example, the mass of the outgoing light particle $d$ is nearly two times of the target $p$, but the momentum that this particle must carry in CM frame is
nearly equal to the whole CM momentum. Hence,
this vector $v^\textrm{l}_\textrm{cm}$ is nearly half the length of $v_\textrm{cm}$. Of course, the precise value depends upon the reaction $Q$-value, but the
basic form of the vector diagram is always the same as what is shown in Fig. \ref{kinematicsininverse}(a). As a result,
the light products emit to forward direction, focusing into a cone of angles around 40$^\circ$ relative to the beam direction.
For the outgoing light particles, there will be two energies for each angle, referred to as low-energy and high-energy branches, which are shown as the thick solid and the thin dashed lines in Fig. \ref{kinematicsininverse}(a), respectively.
The low-energy branch (solid curves) of light particles corresponds to the high-energy branch of heavy particles, as well as to the smaller CM angles of heavy particles,  and hence (typically) to the branch with higher DCSs \cite{Catford}. However, it should be noted that the energy of light particles in this branch is very low, which leads to a lot of difficulties in measuring them.

In the case of (b) $d$ ( A, $p$ )B (or $^3$He( A, $d$)B), the outgoing light products are lighter than the target, thus their CM velocities $v^\textrm{l}_\textrm{cm}$ are larger than $v_\textrm{cm}$, so the vector diagram is very different from that in the case (a), see Fig. \ref{kinematicsininverse}(b). It is obvious that the light particles can emit to backward angles and the maximum angle can be up to 180$^\circ$ in lab frame.  Similar to the case (a), there are two branches. The energy of light particles, corresponding to the branch with higher cross sections, is still very low. This indeed brings in lots of challenges in detecting them.

\begin{figure}[htb]
\includegraphics
  [width=0.98\hsize]
  {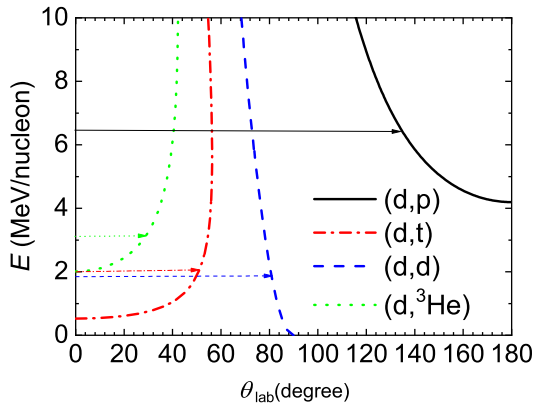}
\caption{Kinematics for different reaction channels induced by a radioactive beam of $^{14}$B at 20 MeV/nucleon impinging on a deuteron target. The arrowed lines point out energies of the light particles (low-energy branch) at the CM angle of 20$^\circ$. }
\label{kinematics}
\end{figure}

In Fig. \ref{kinematics}, kinematics for different reaction channels induced by a radioactive beam of $^{14}$B at 20 MeV/nucleon impinging on a deuteron target is shown. The energies of the outgoing light particles as a function of their angles in lab frame, corresponding to the high cross section part, are given. The dotted, dash-dotted, dashed, and solid curves stand for the reactions of $d$($^{14}$B, $^3$He), $d$($^{14}$B, $t$), $d$($^{14}$B, $d$), and $d$($^{14}$B, $p$), respectively. For each reaction channel, the maximum energy of the light particles, corresponding to a CM angle of 20$^\circ$, is pointed out by the arrowed line. It is obvious that the energies of $^{3}$He, $t$, $d$ and $p$ are less than 3.1, 2.0, 1.9 and 6.5 MeV/nucleon, respectively, which are very low indeed.
They, before being detected, have to punch through the target and loss energies in it. Considering the relatively lower beam intensity and smaller DCSs of transfer reaction (one or two magnitude order lower than elastic scattering), it is better to use a thick target in order to obtain enough statistics for angular distributions. However, considering the energy loss of the light particles in the target and the measurement of light particles, we would like to apply a thin target to precisely detect light particles to get $Q$-value spectrum with better resolution. Therefore, the thickness of target should be specially designed for each experiment in inverse kinematics. For the plastic CH$_2$ or the deuterated polyethylene CD$_2$, the typical thickness is from several hundreds $\mu$g/cm$^2$ to several mg/cm$^2$, see more details in Tab. 2 in Ref. \cite{Wimmer}.

 It is worth noting that the angular scope of light particles produced from different reaction channels is very different. Thus, if the solid angles of detectors are nearly 4$\pi$, several reaction channels can be measured in one experiment. In a (${d}$, ${p}$) reaction in inverse kinematics, the light particles will emit to the backward angles in laboratory frame with the highest cross sections. At the backward angles, protons are almost the only possible products, which makes the particle identification (PID) relatively easier. In a (${d}$, ${t}$) or a (${d}$, $^{3}$He) reaction, the recoil $t$ or $^{3}$He will go to the forward angles, where many particles from other reaction channels can come, so the PID is more important and difficult. In this case, coincidence measurements between the light particles and the residual nuclei are often required.

\subsection{Missing mass method}
For the transfer reaction A(a, b)B in inverse kinematics, the whole CM angular range of the residual particles B can be covered in a small range in laboratory frame. Thus, a ${4\pi}$ coverage for the reaction is easily achieved by placing detectors covering a small range around the beam direction. However, also due to the small range in laboratory system, the energy and angle resolution required for this kind of measurement are too high to be achieved for most real cases. An alternative solution is just to identify the residual particles b, without measuring the residual nuclei B in the forward angle. Therefore, the energies and angles of light particles b are usually measured, and then the excitation energy (or $Q$-value) spectrum of heavy particles B is reconstructed using the functions as follows \cite{Alexandre}.
\begin{equation}\label{qequation}
Q= (\frac{m_a}{m_\textrm{B}}-1)\times E_a+(\frac{m_b}{m_\textrm{B}}+1)\times E_b-\frac{2(m_am_bE_aE_b)^{1/2}\textrm{cos}\theta}{m_\textrm{B}},
\end{equation}
where $m_a$ ($E_a$), $m_b$ ($E_b$), and $m_\textrm{B}$ are the mass (energy) of projectile a, outgoing light particle b, and heavy particle B, respectively. $\theta$ is the outgoing angle of b relative to the beam direction in lab frame.
Although the nucleus of interested $B$ is not measured, all bound and unbound states can be derived with Eq. (\ref{qequation}). Like this, the technique of reconstructing the $Q$-value spectrum, or the excitation energy spectrum ($E_x$), of one of the two ejectiles without measuring it is called the missing mass (MM) method.
The MM technique is one of the few possible techniques for the spectroscopic study of unbound states \cite{Alexandre}. This MM technique is also the most commonly used one for the single-nucleon transfer reaction with a radioactive beam in inverse kinematics. The precise measurement of light particles, including their energies and angles, is the most important thing when the MM method is adopted in experiment.

\begin{figure}[htb]
\includegraphics
  [width=0.9\hsize]
  {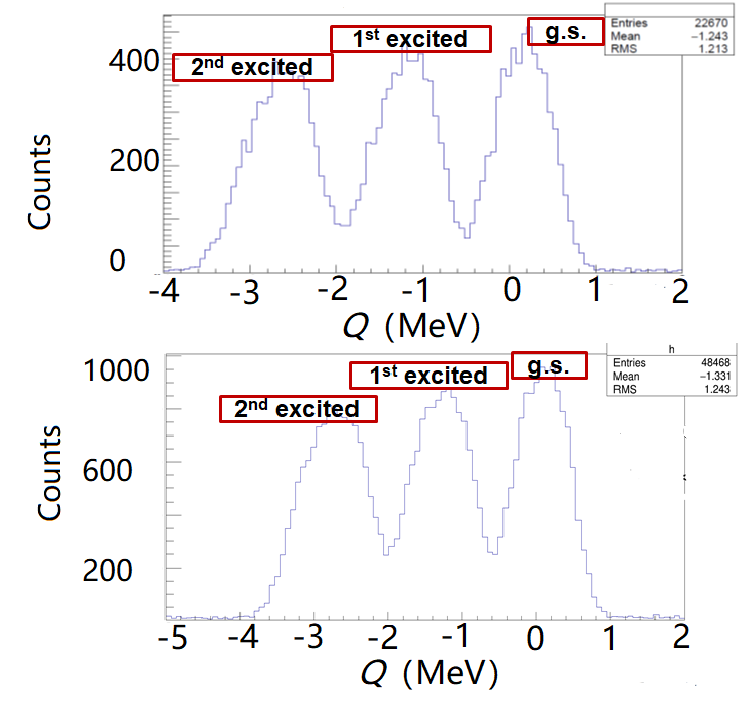}
\caption{The $Q$-value spectra, which were simulated with the Geant4 package \cite{Geant4} for $d$($^{14}$B, $p$) to the ground state, the first ($E_x$ = 1.33 MeV) and the second ($E_x$ = 2.73 MeV) excited states in $^{15}$B at 20 MeV/nucleon. The CD$_2$ target thickness (angular resolution) is 3.0 and 3.6 mg/cm$^2$ (1.8 $^\circ$ and 1.5 $^\circ$) for the upper and lower spectrum, respectively. }
\label{Qsimulation}
\end{figure}

In Fig. \ref{Qsimulation}, the typical $Q$-value spectra reconstructed from the energies and angles of the recoil light particles using the MM method are shown. The spectra were simulated using the GEANT4 package \cite{Geant4},
taking into consideration the beam profile (double gaussian distributions in a circle with a radius of 10 mm), the beam dispersion of radioactive beam (2$\%$ ), the target thickness, the energy threshold (1 MeV), the energy (1$\%$) and angular resolution of light-particle detectors, the energy loss in the target and in the dead layer of light-particle detectors. The simulation is for the reaction of $d$($^{14}$B, $p$) to the ground state, the first ($E_x$ = 1.33 MeV) and the second ($E_x$ = 2.73 MeV) excited states in $^{15}$B with a radioactive beam of $^{14}$B at 20 MeV/nucleon. If the target thickness and angular resolution is 3.0 mg/cm$^2$ and 1.8$^\circ$, respectively, the typical resolution is less than 1 MeV (FWHM). The result is shown in the upper picture of Fig. \ref{Qsimulation}, where three bound excited states in $^{15}$B can be clearly discriminated. It means that if we just measure the energies and the angles of the recoil protons, the resolution of the $Q$-value spectrum is enough to identify different low-lying excited states in $^{15}$B. If we increase the target thickness to 3.6 mg/cm$^2$ and change the angular resolution to 1.5$^\circ$, the corresponding $Q$-value spectrum is shown as the lower one in Fig. \ref{Qsimulation}. Although the resolution is worse than the upper one, the three peaks can still be identified from each other clearly. It means if we increase the target thickness, the $Q$-value resolution will become worse which can be compensated by improving the angular resolution.

In some more complicated cases, if the excitation energies of several states in the unmeasured nucleus are closer to each other, it is hard to discriminate these states just using the MM method. Several methods are often applied to solve this problem. The simplest one is to decrease the target thickness to several hundreds or tens $\mu$g/cm$^2$. However, the statistics often become another new problem if we use this simple method due to the limited beam intensity of the radioactive beam and relatively small cross sections of transfer reactions. Another method is to keep (or increase) the target thickness but add new high-resolution detectors, such as $\gamma$-ray detectors. It is worth noting that the efficiency of $\gamma$-ray is usually very low, the coincidence efficiency of $\gamma$-rays + recoil light particles should be carefully taken into consideration. Finally, the most advanced one is to use the active-target time project chamber (AT-TPC) detector. These will be discussed in details in section. \ref{detector}.

\section{Typical experimental setup}\label{detector}

As stated above, the most important thing is to precisely measure the recoil light particles in order to build a high-resolution excitation-energy (or $Q$-value) spectrum for the unmeasured nucleus $B$. For this purpose, a lot of different detection arrays were constructed around the world. In this section, the frequently used detector setups in different radioactive beam facilities around world for single-nucleon transfer reactions in reverse kinematics are introduced.

\subsection{Silicon detector arrays and ${\gamma}$ detector arrays}
 The energy and angle of the emitting light particles are often measured by the silicon detector arrays. Normally the first layer of the detector arrays made of highly segmented silicon detectors is called ${\Delta}E$ detector, and the light particles can punch through it with a certain energy loss. The second or the third layer is used to stop the light particles, and is called the ${E}$ detector. The large area silicon detectors or CsI(Tl) crystals are usually used. The particle identification (PID) is achieved by the energy loss in ${\Delta}E$ and ${E}$ detector, while the position or angular information is provided by the segmented silicon detector. Sometimes, the energy of the emitting particles is too low to penetrate the ${\Delta}E$ detector. In this case, the time-of-flight (TOF) and ${\Delta}E$ method is applied to identify the recoil light particles, as used in Ref. \cite{Chen-PLB}.

\begin{figure}[!htb]
\includegraphics
  [width=0.9\hsize]
  {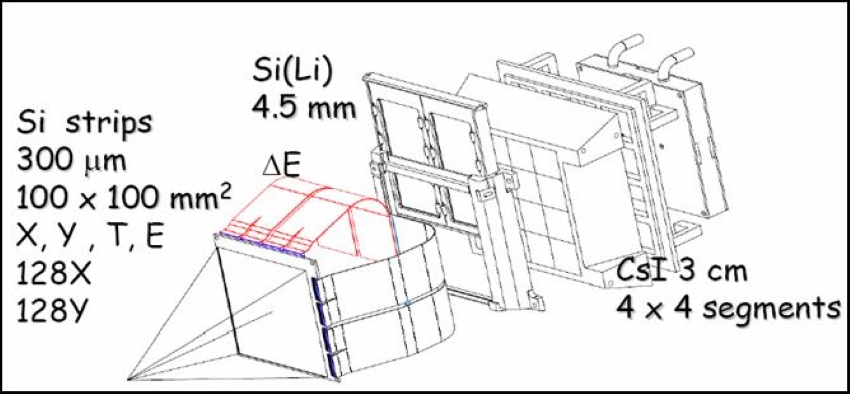}
\caption{Exploded view of the MUST2 telescope. This figure is from Ref. \cite{bib:4}.}
\label{fig:MUST2-module-fig}
\end{figure}

For the past three decades, many silicon detector arrays have been constructed, such as MUST2 \cite{Pollacco-2005} and TIARA \cite{Labiche-2010} at GANIL. MUST2 is a telescope array designed for the detection of the light charged particles produced in direct reactions using the exotic ion beams, with an active area of \SI{10}{\centi\metre} by \SI{10}{\centi\metre} in each module. A typical module is composed of the first stage of double sided silicon-strip detector with 128 strips in each side, followed by the second stage of \SI{4.5}{\milli\metre}-thick Si (Li) segmented with 2 pads of  2 $\times$ 4, and finally a \SI{3}{\centi\metre}-thick CsI segmented into 4 $\times$ 4 pads. The structure of one module is presented in Fig.~\ref{fig:MUST2-module-fig}.

After several times of extension, now the MUST2 array has 10 modules, providing a large angular coverage with efficiency of approximately 70\% up to angles of 45$^\circ$ \cite{Pollacco-2005} by reasonable placement. The combination of hundreds micron meter thick silicon detector and several centimeter thick CsI allows the measurement of large energy range, and at the same time the measurement of time and position. Therefore, the reconstruction of the TOF, momentum, total kinetic energy and trajectory is possible and the construction of the MM spectrum is available. Many experiments have been performed with the help of the MUST2 array, exploring dozens of radioactive isotopes, such as ${\rm ^{9}{He}}$ \cite{bib:5}, ${\rm ^{10}{He}}$ \cite{bib:6}, ${\rm ^{13}{O}}$ \cite{bib:7}, ${\rm ^{21}{O}}$ \cite{bib:8}, and ${\rm ^{61}{Fe}}$ \cite{bib:9}.

\begin{figure}[!htb]
\includegraphics
  [width=0.9\hsize]
  {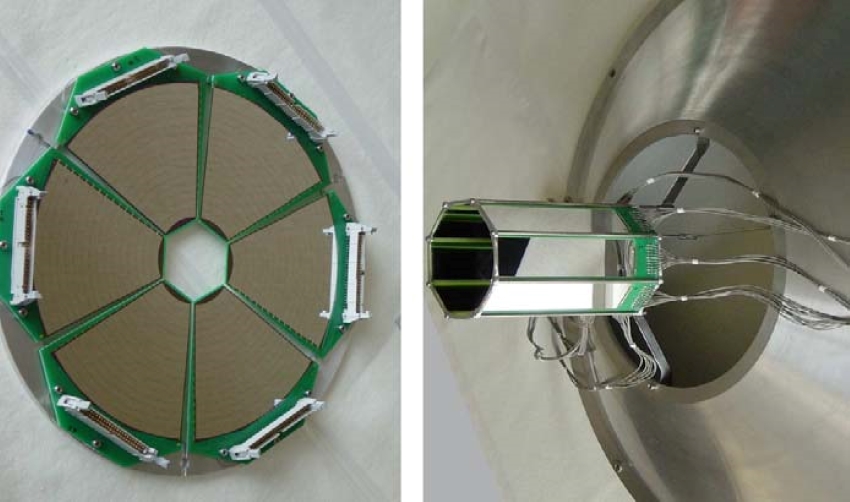}
\caption{The SiHyBall annular detector(left) and the octagonal barrel(right). This figure is from Ref. \cite{Labiche-2010}.}
\label{fig:TIARA-fig}
\end{figure}

In order to cover large scope of angles, a well organized, quasi-4$\pi$ position sensitive silicon array, TIARA, was developed at GANIL. The ultimate goal of TIARA is to perform direct nuclear reaction studies in inverse kinematics using radioactive ion beams. This array is made up of a set of single-layer silicon detectors. The main part consists of a octagonal barrel formed by 8 resistive charge division detectors and a pair of large annular double sided silicon strip detector (SiHyBall) covering each end of the barrel (Fig.~\ref{fig:TIARA-fig}). Considering that the target is almost totally covered by the silicon detectors, unique target changing mechanism, as well as electronics and data acquisition system were developed. Detailed information can be obtained in Ref. \cite{Labiche-2010}.

Despite the good detection performances of silicon detectors, sometimes it is necessary to detect $\gamma$-rays in coincidence for better determination of energy levels or distinguishing long-lived isomeric states. For example, in a $d$(${\rm ^{34}{Si}}$, $p$)${\rm ^{35}{Si}}$ experiment \cite{bib:10} performed at GANIL, energies and angles of the recoil protons were measured with four modules of the MUST2 array, while a more accurate energy determination of bound excitation levels populated in ${\rm ^{35}{Si}}$ is achieved by analyzing the $\gamma$-ray energy spectrum measured by four segmented high purity Germanium (HpGe)  detectors from the EXOGAM array. As for distinguishing long-lived isomeric states, an isomer-tagging technique was used to directly measure the cross section for the $0^+_2$ state in ${\rm ^{12}{Be}}$ populated by the reaction of $d$(${\rm ^{11}{Be}}$, $p$) \cite{Chen-PLB} , and also the delayed-correlation technique was employed in the study of \si{\micro\second}-isomers of ${\rm ^{67}{Ni}}$ \cite{bib:19}. The detection of $\gamma$-ray plays a key role in many single-nucleon transfer reactions \cite{Chen-PLB, bib:12, bib:19}.

\begin{figure}[!htb]
\includegraphics
  [width=0.9\hsize]
  {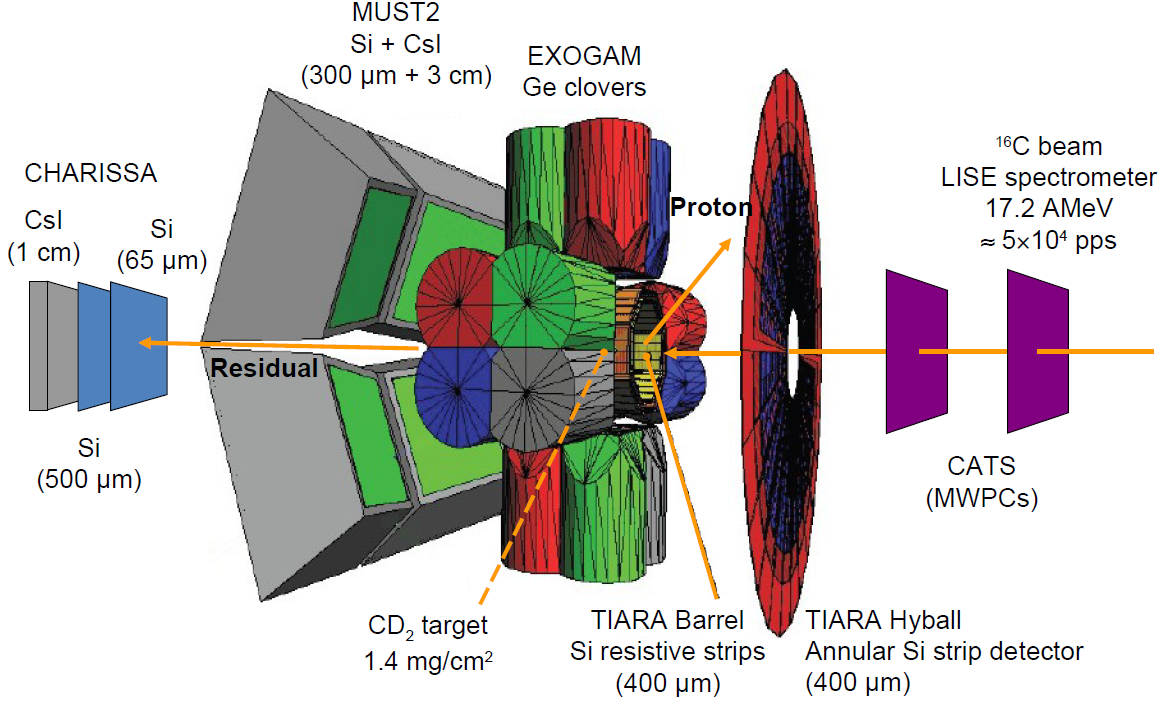}
\caption{A combination view of the detector arrays at GANIL. This figure is from Ref. \cite{bib:51}.}
\label{fig:ganil-combination-fig}
\end{figure}

A combination view of the silicon and high-purity Germanium detection array at GANIL is shown in Fig.~\ref{fig:ganil-combination-fig}. This setup is for the $d$(${\rm ^{16}{C}}$, $p$)${\rm ^{17}{C}}$ experiment \cite{bib:51}.

\begin{figure}[!htb]
\includegraphics
  [width=0.7\hsize]
  {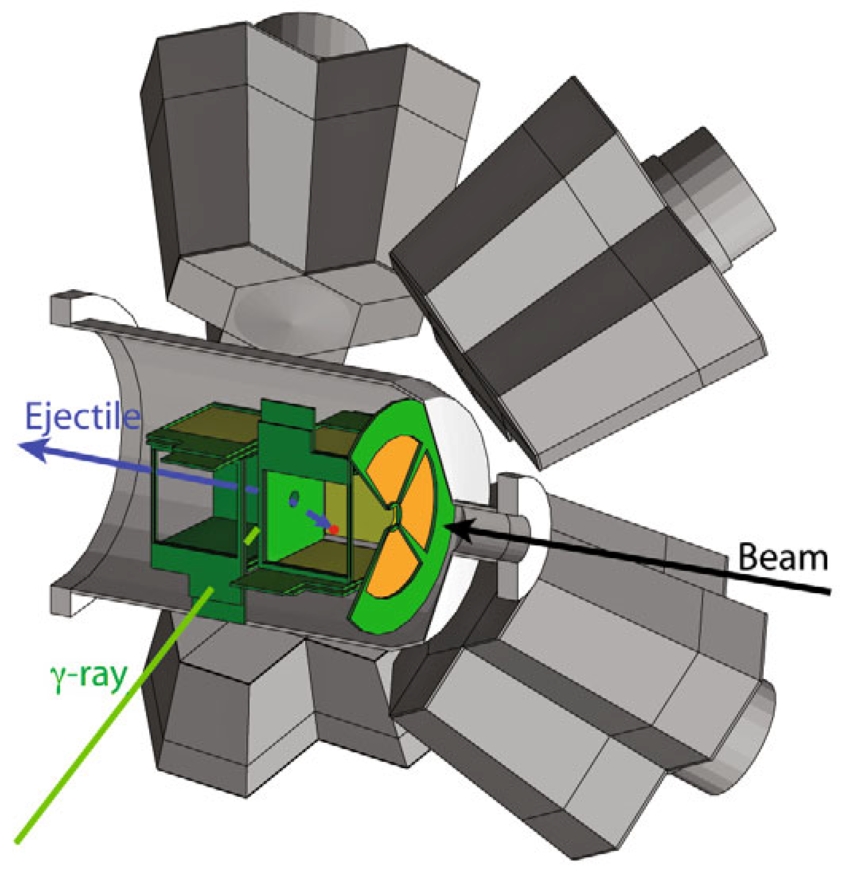}
\caption{The schematic layout of T-REX and Miniball. The left-hand side of the MINIBALL array, vacuum chamber, and particle detectors on is cut away for visualization purposes. This figure is from Ref. \cite{bib:20}.}
\label{fig:trex-miniball-fig}
\end{figure}

The similar arrays were also constructed at other facilities. By using the combination of silicon detectors and $\gamma$-ray detectors, several $(d$, $p)$ transfer reactions in inverse kinematics were performed at REX-ISOLDE, CERN, for the purpose of studying the single-particle properties at the border of the island of inversion \cite{bib:23} or around traditional magic numbers. For instance, in the ${\rm ^{79}{Zn}}$ \cite{bib:18} and ${\rm ^{67}{Ni}}$ \cite{bib:19} experiments, the setup (see Fig.~\ref{fig:trex-miniball-fig}) composed of T-REX array \cite{bib:20} and Miniball \cite{bib:21} allowed the combined detection of protons recoil from the ($d$, $p$) reaction, and of $\gamma$-rays emitting from the residual nuclei. The high-resolution Miniball, which consists of 24 six-fold segmented HpGe crystals, has been used at REX-ISOLDE for over ten years.  An overview of the technical details of the full Miniball setup is given in Ref. \cite{bib:22}. The silicon detector array T-REX was designed to be used in combination with Miniball, providing the positions (or angles) and the ${\Delta E-E}$ PID of light particles.

\begin{figure}[!htb]
\includegraphics
  [width=0.7\hsize]
  {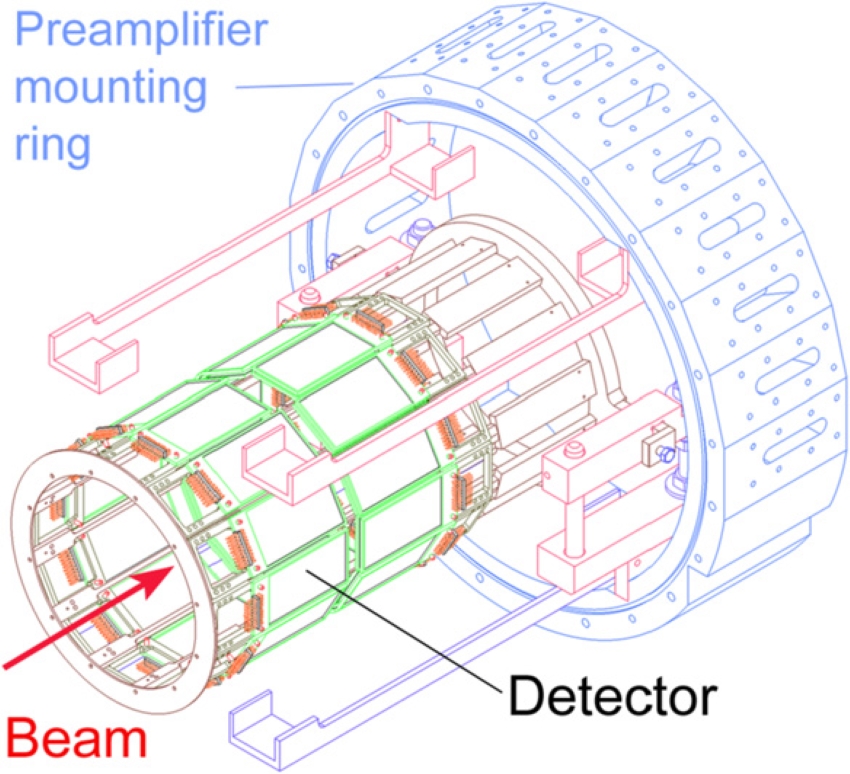}
\caption{The schematic view of ORRUBA. The assembly is mounted on an annular chamber. This figure is from Ref. \cite{bib:30}.}
\label{fig:ORRUBA-fig}
\end{figure}

The silicon detector array developed at HRIBF at Oak Ridge National Laboratory is named ORRUBA \cite{bib:29}, and SuperORRUBA \cite{bib:30}. The latter one is upgraded from the former one. With almost the same geometry, the main difference between these two arrays is that the former one makes use of position sensitive resistive silicon strip detectors, just like TIARA and T-REX, while the upgraded one applies double-sided silicon strip detectors, which have better energy (and position) resolution with the increase of electronics channels needed for the signal readout. The schematic view of ORRUBA is shown in Fig.~\ref{fig:ORRUBA-fig}. Focusing on the evolution of nuclear structure away from the stability line, and the astrophysical $r$-process in supernova, the capability of measuring transfer reactions in inverse kinematics on unstable nuclei has been tested by many experiments \cite{bib:31, Schmitt, bib:33}. Both of them can be used in couple with GAMMASPHERE \cite{bib:34} or GRETINA \cite{bib:35} to obtain the high-resolution excitation energy spectra.

\begin{figure}[!htb]
\includegraphics
  [width=0.8\hsize]
  {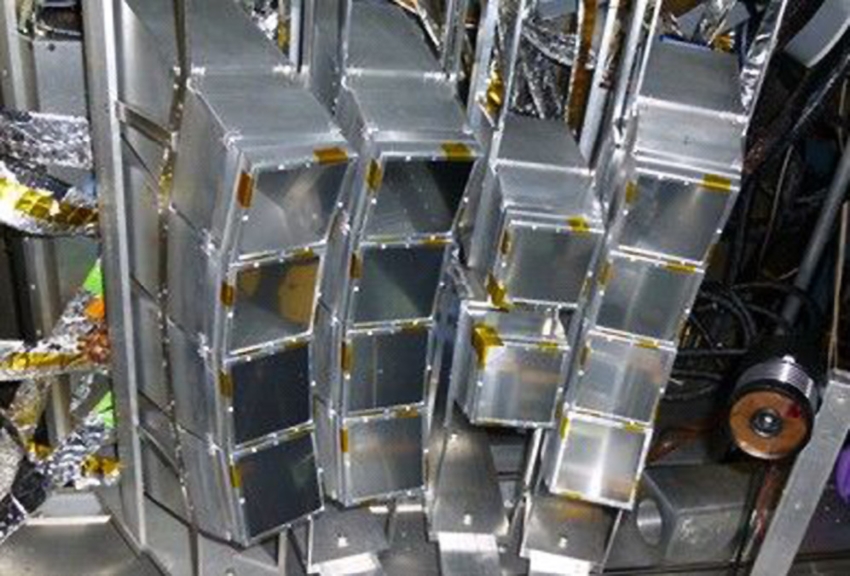}
\caption{The High Resolution Array (HiRA). In this figure, HiRA is placed at the forward angles for a measurement of transfer reaction. This figure is from Ref. \cite{bib:59}.}
\label{fig:hira-fig}
\end{figure}

Almost at the same time, with the construction of ORRUBA, the High Resolution Array (HiRA)\cite{bib:36} was developed at NSCL. Similar to LASSA \cite{bib:37}, the HiRA array uses the combination of silicon detectors and CsI(Tl) crystals for PID. Like MUST2, HiRA is not designed specifically for transfer reactions in inverse kinematics. But with good energy and angular resolution as well as large acceptance, the HiRA array remains to be useful in measuring the transfer reactions. For example, in the study of ${\rm ^{33}{Ar}}$ \cite{bib:38} and ${\rm ^{55}{Ni}}$ \cite{bib:39}, by coincidentally detecting the residual with S800 spectrometer, the excitation energy spectrum was successfully reconstructed from the recoil light particles measured by the HiRA array. Fig.~\ref{fig:hira-fig} shows a picture of HiRA used in a transfer reaction.

\begin{figure}[!htb]
\includegraphics
  [width=0.9\hsize]
  {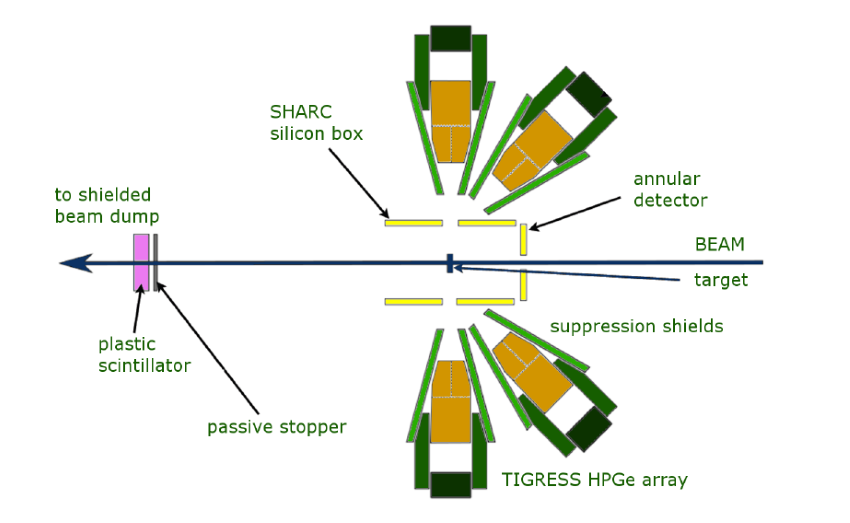}
\caption{Schematic view of the experimental setup composed of the silicon detection array SHARC and the $\gamma$-ray detection array TIGRESS. This figure is from Ref. \cite{Catford}.}
\label{fig:sharc-fig}
\end{figure}

Similar to TIARA, ORRUBA and T-REX, SHARC \cite{bib:41} is a silicon detector array used at TRIUMF for the transfer reaction.  SHARC can be used in conjunction with a $\gamma$-ray detector array TIGRESS \cite{bib:42}, as shown in Fig.~\ref{fig:sharc-fig}. With highly segmented silicon detectors, SHARC provides better angular resolution than the other three arrays. In order to perform transfer reactions and inelastic scattering of rare isotopes in inverse kinematics, a charged particle reaction spectroscopy station IRIS \cite{bib:43} was also developed at TRIUMF. Besides the regular silicon detectors and CsI crystals, IRIS provides a thin solid hydrogen/deuteron target formed by freezing the hydrogen/deuteron gas onto a Ag foil, which was cooled to \SI{4}{\kelvin}.

\begin{figure}[!htb]
\includegraphics
  [width=0.9\hsize]
  {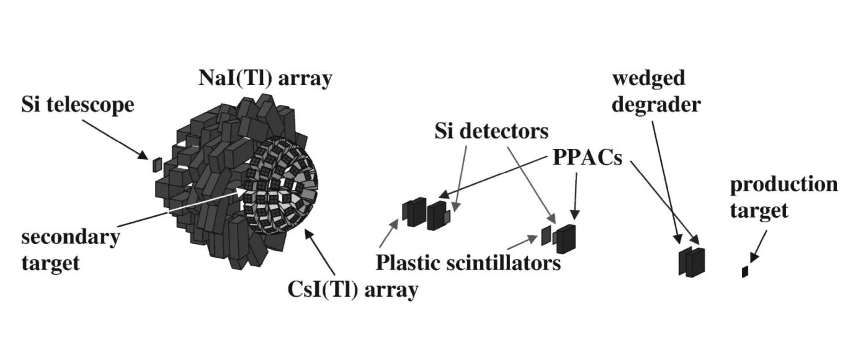}
\caption{Schematic view of the experimental setup in RIKEN for the $d$(${\rm ^{22}{O}}$, $p$) reaction. This figure is from Ref. \cite{bib:61}.}
\label{fig:dali-fig}
\end{figure}

The detection of $\gamma$-rays discussed above was achieved by using HpGe detectors. The HpGe detectors have good energy resolution but low intrinsic detection efficiency, which leads to low statistics in the coincidence measurement of the recoil particles and $\gamma$-rays. There are usually two other methods to solve this problem except increasing the beam intensity. One is to increase the $\gamma$-ray detection efficiency by using other kinds of detectors with higher efficiency, such as NaI. This method was employed by RIKEN for the study of drip-line nuclei. Although RIKEN can provide some radioactive beams with the highest beam intensity in the world, the statistic is still the most difficult question. For example, in a study of the near drip line nucleus ${\rm ^{23}{O}}$ with $d$(${\rm ^{22}{O}}$, $p$) reaction \cite{bib:61} and $d$(${\rm ^{22}{O}}$, $d\gamma$) reaction \cite{bib:74}, the total intensity of the secondary beam was only approximately 1500 counts per second (cps), in which an average intensity of ${\rm ^{22}{O}}$ was 600 cps. It is difficult to provide enough statistics if we make use of the HpGe detector, so a $\gamma$-ray detection array DALI2 comprised of NaI crystals was employed. The experimental setup is shown in Fig.~\ref{fig:dali-fig}. The residuals were analysed by RIPS, while the recoil light particles were measured by 156 CsI(Tl) scintillation crystals. DALI2 \cite{bib:44}, with \SI{20}{\percent} full-energy photon peak efficiency for \SI{1}{\mega\electronvolt} $\gamma$-rays, was placed surrounding the target to detect the $\gamma$-rays from the excited states of ${\rm ^{22}{O}}$ from inelastic scattering \cite{bib:74}. The MUST2 array was also used at RIKEN to detect and identify the recoil particles, such as the $d$($^{11}$Li, ${\rm ^{3}{He}}$) experiment published in Ref. \cite{bib:6}.

\begin{figure}[!htb]
\includegraphics
  [width=0.7\hsize]
  {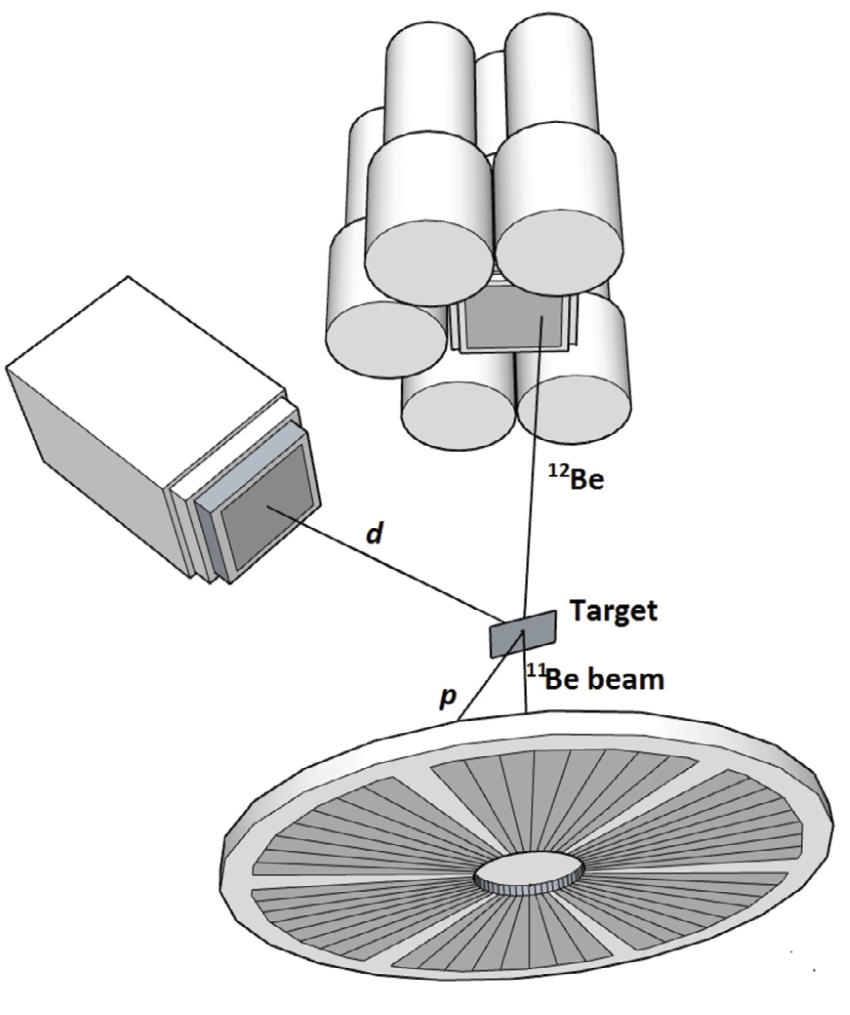}
\caption{Schematic view of the experimental setup for $d$(${\rm ^{11}{Be}}$, $p$) ${\rm ^{12}{Be}}$ performed at RCNP. This figure is from Ref. \cite{Chen-PLB}.}
\label{fig:rcnp-12be-setup-fig}
\end{figure}

There are not any detection arrays at the EN-course terminal at the Research Center of Nuclear Physics (RCNP) in Osaka university \cite{Shimoda}. Therefore, it is an ideal place for users to build their own detection systems for different physical goals using silicon detectors and $\gamma$-ray detectors. Fig.~\ref{fig:rcnp-12be-setup-fig} shows the experimental setup inside the large scattering chamber for the $d$(${\rm ^{11}{Be}}$, $p$) ${\rm ^{12}{Be}}$ experiment at 26.9 MeV/nucleon \cite{Chen-PLB}. The telescopes of TELE0 and TELE1 comprised of silicon detectors and CsI crystals, were used for the detection of the residual nuclei and the scattered light particles, respectively. The annular double-sided silicon strip detector (ADSSD) is responsible for the measurement of recoil protons produced in transfer reaction, and the PID is achieved by the TOF-${\Delta}E$ method. The Scintillation Counter, composed of NaI and BgO scintillator, was used to discriminate the isomeric state in ${\rm ^{12}{Be}}$ from other bound excited states. The similar setup was also applied in $d$(${\rm ^{14}{B}}$, $p$) ${\rm ^{15}{B}}$ experiment to study the $s$-wave intruder components in the low-lying states in ${\rm ^{15}{B}}$. This experimental setup was also employed for the $d$(${\rm ^{16}{C}}$, $^3$He) ${\rm ^{15}{B}}$ and $d$(${\rm ^{15}{C}}$, $^3$He) ${\rm ^{14}{B}}$ experiments, which were performed at the radioactive beam line at Lanzhou (RIBLL1) in 2018 and 2019, respectively.

\subsection{Active target time projection chamber}
With the improvement of the detection techniques used to measure the recoil particles, the major contribution to the uncertainty of the excitation energy spectrum comes from the energy loss inside the target. For the exotic nuclei far from the $\beta$ stability line, the low intensity and low energy features of the secondary beam make it difficult to perform the transfer reaction experiments with traditional plastic (CH$_2$) or deuterated polythene (CD$_2$) target. The combination of time projection chamber and gaseous active target provides an alternative solution for studying the weakly bound nuclear systems \cite{bib:48}.

\begin{figure}[!htb]
\includegraphics
  [width=0.9\hsize]
  {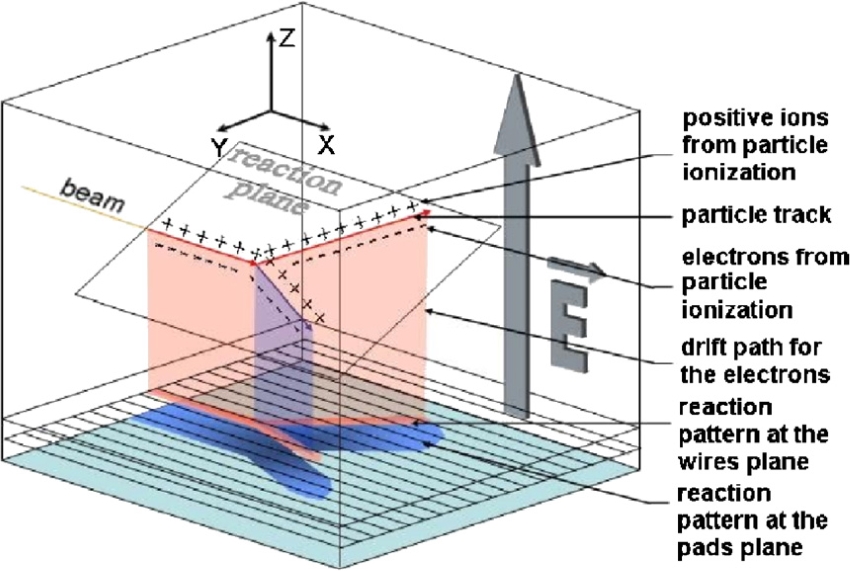}
\caption{Schematic diagram of the ionization process and the detection principle of MAYA. This figure is from Ref. \cite{bib:13}.}
\label{fig:maya-fig}
\end{figure}

The MAYA detector \cite{bib:13}, based on the concept of active target was developed at GANIL more than ten years ago, allowing the use of a relatively thick gaseous target without loss of resolution by using the detection gas as target material. The charged particles inside the detector ionize the filling gas along their trajectories and the released electrons drift toward the amplification area under a high electric field. The projection of the trajectory on one plane is obtained from the segmentation of the readout device, while the third dimension is derived from the measured drift time. The reconstruction of three-dimension trajectories becomes possible by the analysis of pad signals and drift time. Then the reaction point can be derived. Fig.~\ref{fig:maya-fig} shows the ionization process and detection principle of MAYA.

At the same time, this kind of active target time projection chamber (AT-TPC) is a powerful tool to study elastic and inelastic scattering \cite{bib:14, bib:15, bib:16, bib:17}, without worrying about that the low energy scattered particles are stopped inside the target. For the purpose of expanding the dynamical range, lowering thresholds, increasing the detection efficiency, a new-generation active target device, ACTAR TPC \cite{bib:46, bib:47} is under construction at GANIL.

\begin{figure}[!htb]
\includegraphics
  [width=0.8\hsize]
  {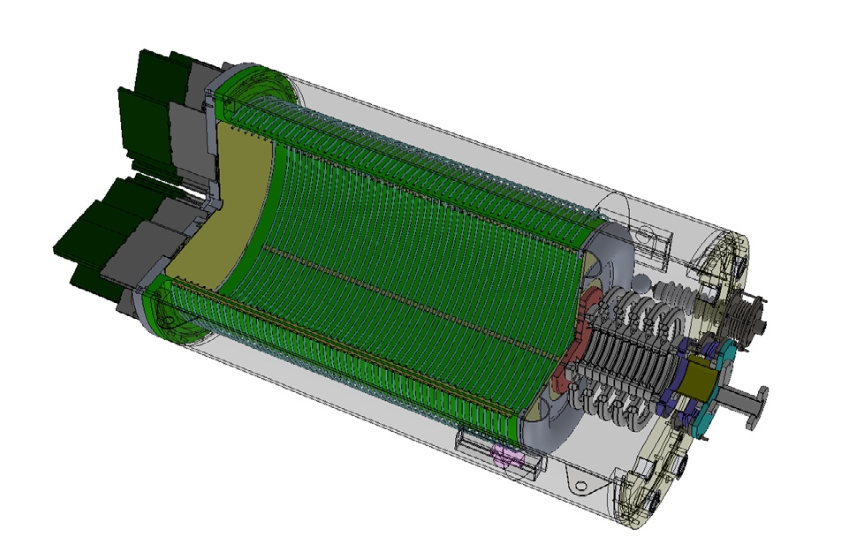}
\caption{A schematic view of the AT-TPC at NSCL. The outer shielding volume was made transparent in this image to make the details of the inner volume more visible. Beam enters the detector through the beam duct at the right-hand side of the image and moves toward the sensor plane on the left. This figure is from Ref. \cite{bib:55}.}
\label{fig:nscl-attpc-fig}
\end{figure}

Other facilities are also making progresses in this state-of-the-art detector. The larger AT-TPC constructed at NSCL \cite{bib:56, bib:55, bib:40} can be used to measure longer trajectories of recoil particles. The test experiment with a ${\rm ^{46}{Ar}}$ beam shows the potential of this detector in investigating the single-particle states of nuclei far from the stability line. Fig.~\ref{fig:nscl-attpc-fig} shows the schematic view of the AT-TPC at NSCL.

Following the frontier, a compact AT-TPC, MAIKo, has been developed at RCNP, together with Kyoto University \cite{bib:49}. The elastic and inelastic scattering experiments of a radioactive beam of ${\rm ^{10}{C}}$ \cite{bib:50} on ${\rm ^{4}{He}}$ have been performed to test the performance of MAIKo \cite{bib:49}. This kind of advanced detector AT-TPC is also under development by the experimental group at Peking University \cite{bib:57}, Institute of Applied Physics in Shanghai \cite{bib:58}, and Institute of Modern physics in Lanzhou, China. 

\subsection{A new approach with magnetic spectrometer}
Unlike the traditional magnetic spectrometers for measuring reactions in normal kinematics, a new type of helical orbit spectrometer, HELIOS \cite{bib:54, bib:24}, was developed at the Argonne National Laboratory. HELIOS, which was specially designed to measure the recoil light particles with high energy and position resolution, has been used in many transfer reaction experiments in inverse kinematics \cite{bib:25, bib:26, 16C-Wuosmaa, bib:28}.

\begin{figure}[!htb]
\includegraphics
  [width=0.9\hsize]
  {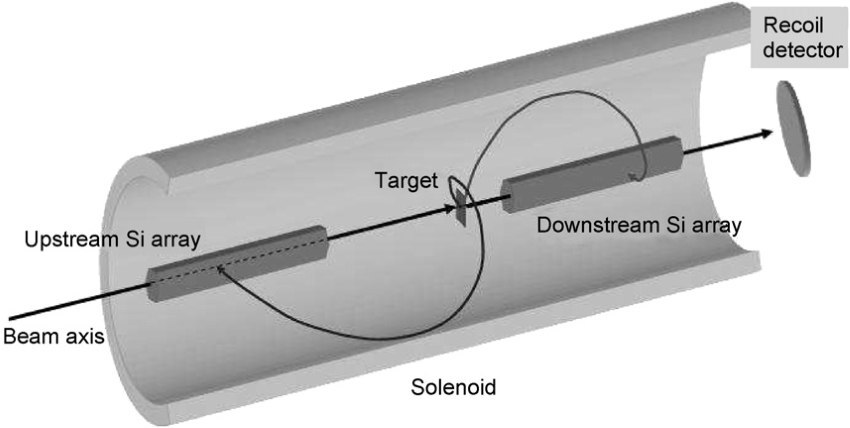}
\caption{HELIOS at the Argonne National Laboratory. This figure is from Ref. \cite{bib:24}.}
\label{fig:helios-scheme-fig}
\end{figure}

By placing the target and detectors inside a uniform magnetic field, the charged particles emitted from the reaction will travel on cyclotron orbits in the magnetic field and reach the beam axis again where they are detected by silicon detectors after one cyclotron period, as shown in Fig.~\ref{fig:helios-scheme-fig}. By measuring the arrival time at the silicon array along the beam axis, which is independent of energy and scattering angle, the mass to charge ratio ${A/q}$ can be obtained, thus the PID can be achieved. For a fixed reaction $Q$-value, the energy of the emitting particle is proportional to the $z$ position measured by the silicon arrays along the beam direction. With smaller influence of the target thickness and the beam spot size, a better resolution for the excitation energy is achieved, as discussed in Ref. \cite{bib:24}. In the $d$(${\rm ^{12}{B}}$, $p$)${\rm ^{13}{B}}$ \cite{bib:28} experiment, the resolution of HELIOS is enough to separate two closely spaced excited states at $E_x$ =  \SI{3.48}{\mega\electronvolt} and \SI{3.68}{\mega\electronvolt} in ${\rm ^{13}{B}}$. Although this method will encounter problems when measuring particles with the same ${A/q}$ value, such as deuterons and $\alpha$ particles, it is enough for most transfer reactions.

Due to the introduction of the magnetic field, the acceptance of the spectrometer is limited by the radius of the solenoid and the homogeneity of the magnetic field strength. Combining with $\gamma$-ray detectors remains a challenge because the light particles will travel inside the volume where the $\gamma$-ray detectors are supposed to be placed.

\section{Experimental results of single-nucleon transfer reactions}\label{experimentalresults}
In most light neutron-rich nuclei around $N$ = 8, the ordering of $2s_{1/2}$, $1p_{1/2}$, and $1d_{5/2}$ orbits is usually different from that in stable nuclei, resulting in the emergence of large amount $s$-wave intruder component and the formation of halos \cite{bib:62, Tanihata}, such as $^{11}$Be, $^{11}$Li and $^{15}$C. The breakdown of the $N$ = 8 shell closure due to the $2s_{1/2}$ orbital intruding into the $p$-shell is closely related to the appearance of the neutron halo in the light neutron-rich isotopes. It is important therefore to study systematically the influence of the $2s_{1/2}$ orbital around $N$ = 8.

Tab.~\ref{tab:n-eight-tab} summarized the intruder $s$-wave intensities in the ground-state wave functions of loosely bound nuclei around $N$ = 8 from various experiments, together with their single-neutron separation energies ($S_{n}$). It is worth noting that the ground states of most nuclei with smaller $S_{n}$ (${S_{n} \leq \SI{1.22}{\mega\electronvolt}}$), such as ${\rm ^{11}{Be}}$ and ${\rm ^{15}{C}}$, have the predominant $s$-wave component, but their neighboring nuclei with one more neutron and larger $S_{n}$ (${S_{n} \ge \SI{3.17}{\mega\electronvolt}}$), ${\rm ^{12}{Be}}$ and ${\rm ^{16}{C}}$, have a little such component. In addition to ${\rm ^{11}{Be}}$ and ${\rm ^{15}{C}}$, ${\rm ^{14}{B}}$ is another nucleus with smaller $S_{n}$ and dominant $s$-wave component \cite{bib:64, bib:65, bib:66, Bedoor, bib:68}, which indicates that the $2s_{1/2}$ orbit intrudes into $1d_{5/2}$ orbital in the ground state of ${\rm ^{14}{B}}$ when compared with the stable nuclei. Therefore, it is an interesting question about how much $s$-wave component in the ground state of its adjacent isotope ${\rm ^{15}{B}}$. Moreover, such an intruder component has been measured for other $N$ = 10 isotones by many experiments, such as 2$n$ removal reaction for ${\rm ^{14}{Be}}$ \cite{bib:69} and 1$n$ transfer reaction for ${\rm ^{16}{C}}$ \cite{16C-Wuosmaa}.

\begin{table}[!htb]
\caption{Summary of intruder $s$-wave strength in the ground state wave functions of weakly bound nuclei around $N$ = 8, together with their single-neutron separation energies ($S_{n}$).}
\label{tab:n-eight-tab}
\begin{tabular*}{8cm} {@{\extracolsep{\fill} } llr}
\toprule
Isotope & $s$-wave strength & $S_{n}$ (\SI{}{\mega\electronvolt}) \\
\midrule
${\rm ^{11}{Li}}$ & \SI{47}{\percent} \cite{kobayashi-1988, bib:71, bib:72} & \SI{0.396}{} \\
${\rm ^{11}{Be}}$ & \SI{71}{\percent} \cite{Schmitt} & \SI{0.504}{} \\
${\rm ^{12}{Be}}$ & \SI{19}{\percent} \cite{Chen-PLB} & \SI{3.17}{}  \\
${\rm ^{14}{Be}}$ & \SI{87}{\percent} \cite{bib:69} & \SI{1.78}{} \\
${\rm ^{13}{B}}$  & little \cite{bib:28, bib:73} & \SI{4.88}{}  \\
${\rm ^{14}{B}}$  & \SI{71}{\percent} to \SI{89}{\percent} \cite{bib:64, bib:65, bib:66, Bedoor, bib:68} & \SI{0.97}{}  \\
${\rm ^{15}{B}}$  & no data & \SI{2.78}{} \\
${\rm ^{15}{C}}$  & \SI{88}{\percent} \cite{bib:79} & \SI{1.22}{}  \\
${\rm ^{16}{C}}$  & \SI{30}{\percent} \cite{16C-Wuosmaa} & \SI{4.25}{} \\
\bottomrule
\end{tabular*}
\end{table}

Transfer reactions, especially the single-nucleon transfer reaction, can provide very useful spectroscopic information in understanding the evolution of nuclear shell structure by precisely detecting the unusual components, such as the intruder $s$-wave. Focusing on the spectroscopic study of neutron-rich He, Li, Be, B and C isotopes (including bound and unbound nuclei) around $N$ = 8, various single-nucleon transfer reaction experiments performed with light exotic nuclei in inverse kinematics will be reviewed in this section. The results obtained from other kinds of reaction, such as breakup reaction, charge exchange reaction, and knock-out reaction will not be presented here.

\subsection{Helium isotopes}

\textbf{$^8$He}

The ``double-borromean" nucleus $^8$He \cite{lemasson-2010} is an interesting subject with the largest neutron-to-proton ratio among all the known particle-stable nuclei, exhibiting a neutron halo or thick neutron skin. Thus, $^8$He is an excellent candidate for the test of different nuclear structure models \cite{keeley-2007}. Its neighbours $^7$He and $^9$He are particle unbound, thus it offers an opportunity to study the shell evolution of nuclear structure as a function of an increasing number of neutrons.
The charge radius of $^8$He is smaller than that of $^6$He due to the more isotropic distribution of the four valence neutrons \cite{Mueller-2007}, which is different from our traditional concept. With four loosely
bound valence neutrons, $^8$He is an unique system for investigating
the role of neutron correlations, such as pairing \cite{lemasson-2009,lemasson-2011}.

The ground state of $^8$He is still an ambiguous topic, attracting continuous attentions experimentally as well as theoretically. The cluster orbital shell model approximation (COSMA) assumes that $^8$He is comprised of a $^4$He core plus four valence neutrons filling the $1p_{3/2}$ sub-shell \cite{zhukov-1994}. This assumption is equal to pure \emph{jj} coupling.
 The consistent analysis of $p$($^8$He, $t$) reaction at incident energies of 15.7 and 61.3 MeV/nucleon \cite{keeley-2007} shows that the ground state wave function of $^8$He deviates from the pure ($1p_{3/2}$)$^4$ structure. This result is in agreement with the theoretical calculation of Hagino's group \cite{hagino-2008}, which concludes that the
probability of the $(1p_{3/2})^4$ and $(1p_{3/2})^2(1p_{1/2})^2$ configurations in the ground state wave function of $^8$He nucleus
is 34.9$\%$ and 23.7$\%$, respectively. The asymmetry molecule dynamics (AMD) calculation also suggests that the ground state
of $^8$He has both the \emph{jj} coupling feature ($^4$He + 4\emph{n}) and the \emph{LS} coupling feature ($^4$He + 2\emph{n}
+ 2\emph{n}) \cite{kanada-2007}. For the ground state of neutron-rich He isotopes, the ``\emph{t} + \emph{t} + valence neutrons" structure \cite{aoyama-2006} and ``di-neutron" structure \cite{itagake-2008} are also predicted by the AMD model.
  The contributions from the first 2$^+$ excited state of $^6$He and the cluster of $^5$H
to the ground state configuration of $^8$He have also been investigated by $p$($^8$He, $t$) transfer reaction at an incoming energy of 25 and 61.3 MeV/nucleon \cite{wolski-2002,korsheninnikov-2003}.
Early in 2001, a heavy hydrogen $^5$H had been observed by \emph{p}($^6$He, 2\emph{p})$^5$H reaction at 32 MeV/nucleon. By using triple 2\emph{p}-\emph{t} coincidences, the $^5$H ground state was identified as a resonance state at 1.7 $\pm$ 0.3 MeV above the \emph{n} + \emph{n} + \emph{t} threshold, with a width of
1.9 $\pm$ 0.4 MeV \cite{Korsheninnikov-2001}.

Transfer reaction is a sensitive and powerful probe of the properties of atomic nuclei \cite{kate-2013}.  The neutron $SF$ extracted from $p$($^8$He, $d$)$^7$He reaction includes the structure information of $^8$He and $^7$He. The ground state of $^7$He is a resonant state which often decays into $^6$He + \emph{n} immediately, while an excited
state of $^7$He at $E_x$ = 2.9 $\pm$ 0.3 MeV mainly decays into
3\emph{n} + $^4$He \cite{korsheninnikov-1999}. Therefore, we often measure the deuterons in coincidence with $^6$He and $^4$He for the reaction channel of $p$ ($^8$He, $d$)$^7$He$_{g.s.}$ and $p$($^8$He, $d$)$^7$He$_{E_{x} = \SI{2.9}{\mega\electronvolt}}$, respectively.
The ratio of 2\emph{n} to 1\emph{n} transfer cross section for $^8$He is expected to be sensitive to the correlations among the valence neutrons \cite{hagino-2008,hagino-2009}.
Translational invariant shell model (TISM) calculations suggest that the ratio of $SF$s for $p$($^8$He, $t$)$^6$He$_{g.s.}$ to $p$($^8$He, $t$)$^6$He$^{2+}$ depends strongly on the assumed structure of $^8$He ground state \cite{wolski-2002,korsheninnikov-2003}. It is worth mentioning that this ratio is quite close to 1.0 for pure \emph{jj} coupling assumption, but deviates from 1.0 for the mixture of \emph{jj} and \emph{LS} coupling.
The $^6$He ground state with a half-time of 806.70 ms can be directly measured, but its 2$^+$ excited state will decay to $^4$He + 2$n$ immediately. Experimentally, the coincidence of \emph{t }+ $^4$He is usually observed for the $p$($^8$He, $t$)$^6$He$^{2+}$ reaction.

\begin{figure}[!htb]
\includegraphics
  [width=1.05\hsize]{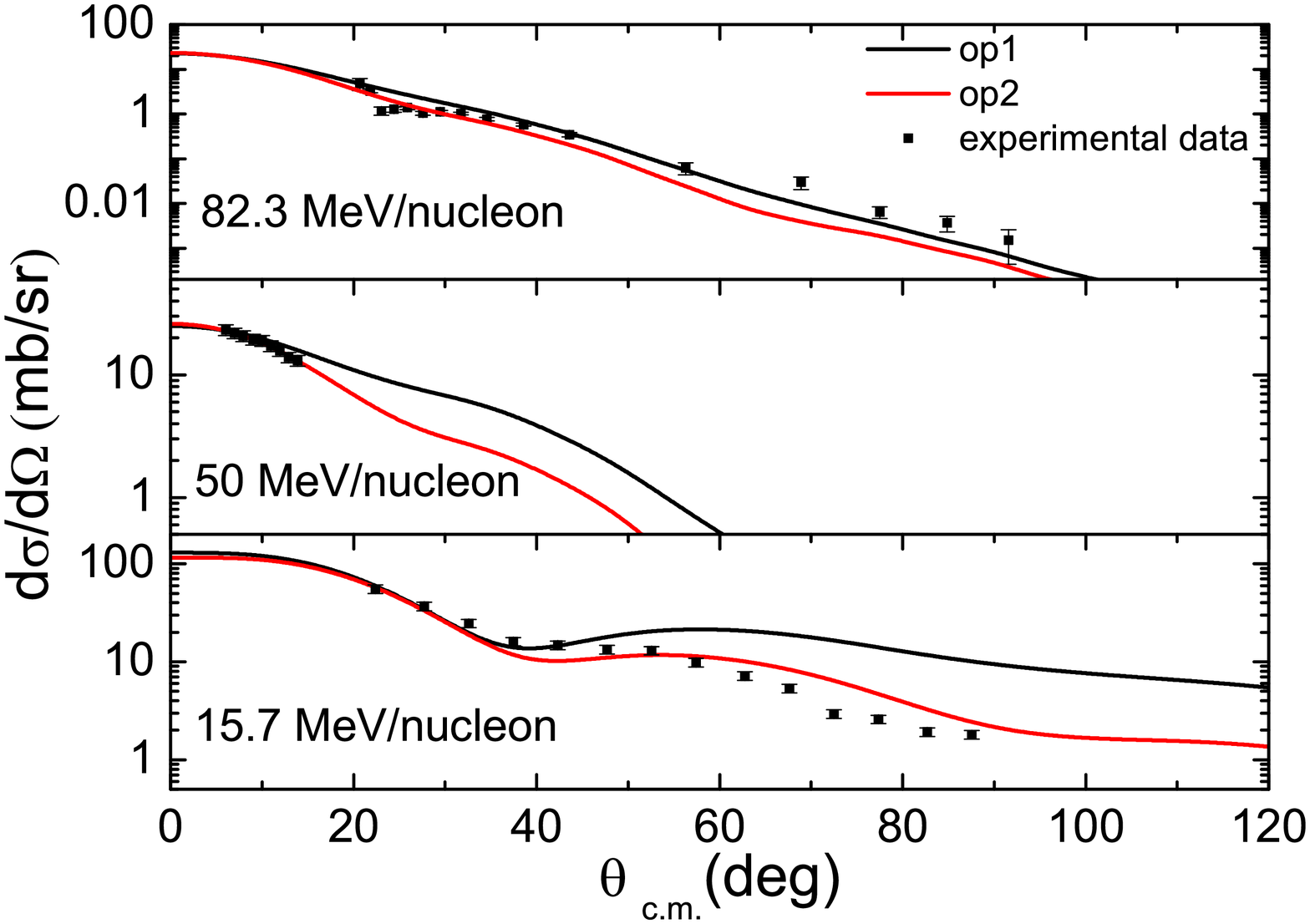}
\caption{Compared with the theoretical calculations, the experimental DCSs of $^8$He($p$, $d$)$^7$He$_{g.s.}$ reaction at incoming energies of 82.3, 50 \cite{korsheninnikov-1999} and 15.7 MeV/nucleon \cite{skaza-2006}.
 The black line represents the initial FR-ADWA calculation result without any parameter adjustment.
 If imaginary part depth of the \emph{d} + $^7$He single-folding potential increases by a factor of 2.0, the results are shown as the red solid lines.}\label{8He-1n-all}
\end{figure}

The DCSs for
reactions $p$($^8$He, $d$)$^7$He$_{g.s}$ and $p$($^8$He, $d$)$^6$He$^{2+}$  were measured at a relatively high incident
energy of 82.3 MeV/nucleon. The experiment was performed at RIKEN by nuclear physics experimental group at Peking University. The results of $p$($^8$He, $d$)$^7$He$_{g.s}$ is shown in Fig. \ref{8He-1n-all}. These results were analyzed in the framework of FR-ADWA \cite{Schmitt} with the code FRESCO \cite{Fresco}. It was found that although the extracted $SF$s are little different owning to the choice of different OP parameters, the values are all dramatically smaller than 4.0.
 This may indicate that the ground state wave function of $^8$He is not pure \emph{jj} coupling or $(1p_{3/2})^4$ configuration, and some other configurations, such as $(1p_{3/2})^2$$(1p_{1/2})^2$ may have some probability.
The $p$($^8$He, $t$) $^6$He$^{2^+}$  channel is prior to the $p$($^8$He, $t$) $^6$He$_{g.s.}$ channel at 82.3 MeV/nucleon, which is also observed at incident energies of  25 \cite{wolski-2002} and 61.3 \cite{korsheninnikov-2003} MeV/nucleon.  For all the existing ($p$, $t$) reaction data with larger angular range in the CM frame,  the consistent analysis shows that the $^5$H cluster transfer is necessary for reproducing the angular distributions at angles larger than 90$^\circ$. Thus, the cluster structure of $^8$He = $^5$H + $^3$H might not be neglected in the ground state of $^8$He. Extracted from all the existing ($p$, $t$) reaction data, all the ratios of $p$($^8$He, $t$)$^6$He$_{g.s.}$ to $p$($^8$He, $t$)$^6$He$^{2+}$ deviate from 1.0, which is inconsistent with the hypothesis of four valences filling a closed 1p$_{3/2}$ sub-shell.

\textbf{$^{9}$He}

As stated above, for the ground state of $^{8}$He, besides four valence neutrons filling the $1p_{3/2}$ shell, some other configurations, such as $(1p_{3/2})^2$$(p_{1/2}$)$^2$, may have some probability. How about the ground state of ${\rm ^{9}{He}}$, one more valence neutron than ${\rm ^{8}{He}}$?
Some theoretical calculations predicted that, with the same neutron number, the ground state of ${\rm ^{9}{He}}$ is particle unbound and may show the same parity inversion observed in the neighboring ${\rm ^{11}{Be}}$ and ${\rm ^{10}{Li}}$ (see the subsection below). However, in a no core shell model calculation \cite{bib:95}, they found that the ${\rm ^{9}{He}}$ ground-state resonance has a negative parity and thus breaks the parity-inversion mechanism found in the ${\rm ^{11}{Be}}$ and ${\rm ^{10}{Li}}$ nuclei.

The unbound nuclear system ${\rm ^{9}{He}}$ was investigated through the $d$(${\rm ^{8}{He}}$, $p$) transfer reaction by M. S. Golovkov \emph{et al} using a ${\rm ^{8}{He}}$ beam \cite{bib:93} at a laboratory energy of $25$ MeV/nucleon. The lowest resonant state of ${\rm ^{9}{He}}$ was found at \SI{2.0}{\mega\electronvolt} with a width of about \SI{2.0}{\mega\electronvolt}, which mainly came from the experimental energy resolution of about \SI{0.8}{\mega\electronvolt}. The observed angular correlation pattern is uniquely explained by the interference of the $(1/2)^{-}$ resonance with a virtual $(1/2)^{+}$ state and with a $(5/2)^{+}$ resonance at energy $\ge \SI{4.2}{\mega\electronvolt}$.
However, the experimental energy resolution of about \SI{800}{\kilo\electronvolt} prevents to conclude in strong disagreement with previous two-proton knockout reaction experiment \cite{bib:53}, which suggests that the ground-state of ${\rm ^{9}{He}}$ has ${J}^{\pi}={1/2}^{+}$.

The same reaction was also performed at GANIL with the ${\rm ^{8}{He}}$ beam \cite{bib:5} at a laboratory energy of 15.4 \SI{}{\mega\electronvolt}/nucleon. The MM spectrum was deduced from the kinetic energies and the emission angles of protons detected by four MUST2 telescopes. This MM spectrum displays a structure just above the neutron emission threshold, which is identified as the ground state of ${\rm ^{9}{He}}$. Despite limited statistics of the angular distribution for the state observed very close to the ${\rm ^{8}{He}}$ + $n$ threshold ($180 \pm 85$ \SI{}{\kilo\electronvolt} above), the result supports the conclusion that the ground state spin-parity of ${\rm ^{9}{He}}$ is $(1/2)^{+}$. This result confirms the parity inversion and $s$-wave intrusion. The first excited state lies approximately \SI{1.3}{\mega\electronvolt} above the neutron threshold and is compatible with $J^{\pi}=(1/2)^{-}$, exhibiting at the same time a strong mixed nature.

\subsection{Lithium isotopes}

\textbf{$^{10}$Li}

The study of the unbound system ${\rm ^{10}{Li}}$ is of great interest since knowledge on this system is a necessary ingredient in a theoretical description of the Borromean nucleus ${\rm ^{11}{Li}}$ \cite{bib:88}. In addition, the existence of a low-lying virtual intruder $s$-state has been predicted by some theoretical models. But due to the unbound nature of ${\rm ^{10}{Li}}$, experimental study of this nuclei encounters a lot of challenges, such as the coincidence detection of the neutron and ${\rm ^{9}{Li}}$.

Two different $d$(${\rm ^{9}{Li}}$, $p$)${\rm ^{10}{Li}}$ transfer reactions in inverse kinematics were performed separately at REX-ISOLDE \cite{bib:89} and NSCL \cite{bib:90}. The results of the former experiment \cite{bib:89} supported the existence of a low-lying $(s)$ virtual state, with a (negative) scattering length and a $p_{1/2}$ resonance with an energy of $E_{r} \simeq \SI{0.38}{\mega\electronvolt}$. However, the results of the latter one \cite{bib:90} were inconclusive as to the possible presence of a low-lying virtual state in ${\rm ^{10}{Li}}$ due to the very poor statistics. Actually the low statistics and the poor energy resolution have always prevented definitive conclusions on the study of the structures of ${\rm ^{10}{Li}}$. The results from these two experiments were reexamined with a joint and consistent analysis \cite{bib:92}. From this analysis, they concluded that both measurements can be described consistently using the same model for the $n-{\rm ^{9}{Li}}$ interaction and the seemingly different features can be understood if we take into consideration of the different incident energy and angular range covered by the two experiments. For the lower beam energy, the $s$-wave virtual state in ${\rm ^{10}{Li}}$ plays a key role, but for the higher energy experiment, the excitation energy below ${{E}_{x} < \SI{1}{\mega\electronvolt}}$ is dominated by the ${p}_{1/2}$ resonance.

\begin{figure}[!htb]
\includegraphics
  [width=0.9\hsize]
  {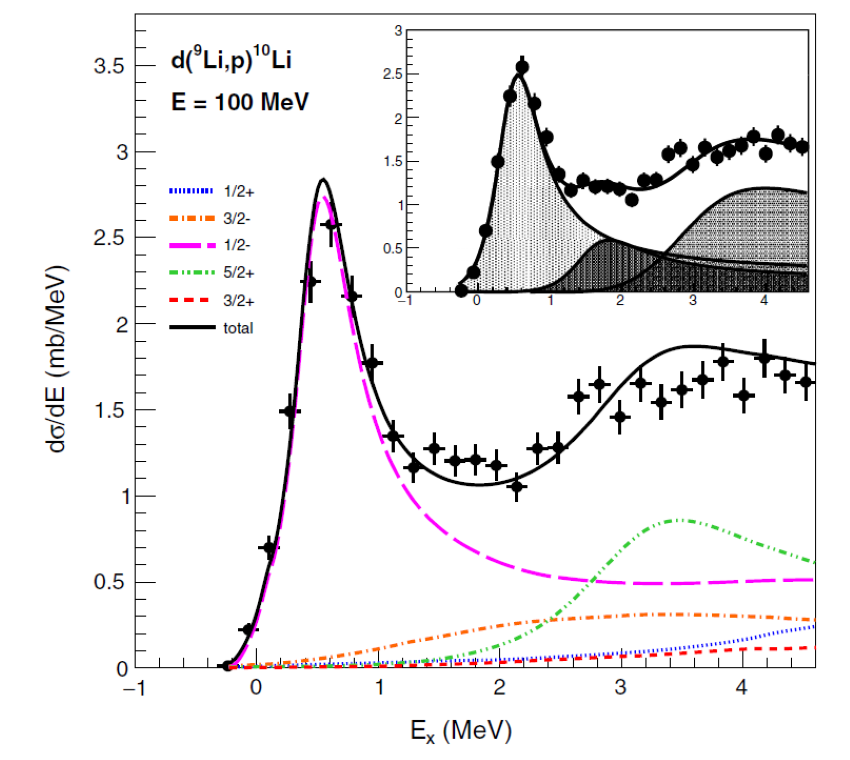}
\caption{${\rm ^{10}{Li}}$ energy spectrum rebuilt from the recent $d$(${\rm ^{9}{Li}}$, $p$)${\rm ^{10}{Li}}$ reaction. The curves show the partial wave contributions obtained from the theoretical predictions, including paring correlation effects. The solid black line represents a sum of the contribution from each partial wave. The inset picture is the best-fitting sum of three Fano functions convoluted with the experimental energy resolution. This figure is from Ref. \cite{bib:87}.}
\label{fig:li10-energy-fig}
\end{figure}

Most recently, a new $d$(${\rm ^{9}{Li}}$, $p$)${\rm ^{10}{Li}}$ transfer reaction in inverse kinematics was performed at TRIUMF with a much higher beam intensity ($10^{6}$ pps), which guaranteed significant statistics \cite{bib:87}. The ${\rm ^{10}{Li}}$  energy spectrum, which was analyzed by using three Fano functions, is shown in Fig.~\ref{fig:li10-energy-fig}. The comparison between experimental data and theoretical predictions, including pairing correlation effects, shows clearly the existence of a $p_{1/2}$ resonance (purple dashed line in Fig.~\ref{fig:li10-energy-fig}) at ${0.45}\pm{0.03}$ \SI{}{\mega\electronvolt} excitation energy. However, there is no obvious evidence for the existence of a significant $s$-wave contribution (blue dotted line) close to the threshold energy.
Moreover, two high-lying resonances at 1.5 and 2.9 \SI{}{\mega\electronvolt} are observed. The corresponding angular distributions indicate a significant $s_{1/2}$ partial wave contribution (blue dotted line) for the 1.5 \SI{}{\mega\electronvolt} resonance and a mixing of configurations at $E_{x}=\SI{2.9}{\mega\electronvolt}$ for the first time, with the $d_{5/2}$ partial-wave (green dot-dashed line) contributing the most to the cross section.

 There are a large number of other kinds of experiments. Although the excitation energy, parity and spin assignments were controversial, one can conclude that the ground state of $^{10}$Li contains a valence neutron in a $2s_{1/2}$ state at approximately 50 keV or below \cite{Tanihata}. This conclusion is different from that obtained from the transfer reactions stated above, which might be attributed to the difficulty of coincidence measurement between $^9$Li and proton for such low-energy resonance using transfer reaction.

\textbf{$^{11}$Li}

\begin{figure}[!htb]
\includegraphics
  [width=0.8\hsize]
  {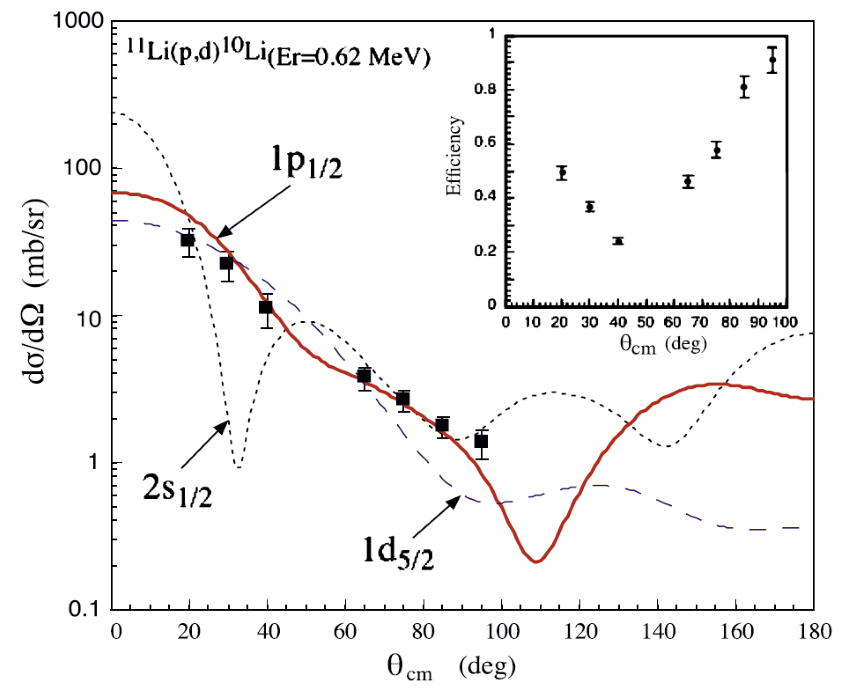}
\caption{The angular distribution for ${\rm ^{11}{Li}}$($p$, $d$)${\rm ^{10}{Li}}_{E_{r}=\SI{0.62}{\mega\electronvolt}}$. The solid (red) curve, dashed (blue) curve, and dotted (black) curve represent the DWBA calculations with an assumption that one neutron is removed from the $1p_{1/2}$, $1d_{5/2}$, and $2s_{1/2}$ orbital, respectively. The detection efficiency is shown in the inset. This figure is from Ref. \cite{bib:91}.}
\label{fig:li10-angular-distributions-fig}
\end{figure}

In order to study the configurations in the ground state of ${\rm ^{11}{Li}}$, a $p$(${\rm ^{11}{Li}}$, $d$)${\rm ^{10}{Li}}$ reaction was performed at TRIUMF using a solid hydrogen target and a 5.7 MeV/nucleon $^{11}$Li beam \cite{bib:91}. Only one resonance at $E_{r} = 0.62 \pm 0.04$ MeV  with a width of $0.33 \pm 0.07$ MeV was observed. The elastic scattering was measured along with the transfer reaction to obtain the OP parameters for the entrance channel. The angular distribution of $p$(${\rm ^{11}{Li}}$, $d$)${\rm ^{10}{Li}}$ to $E_{r} = \SI{0.62}{\mega\electronvolt}$ is best reproduced by assuming the neutron is removed from the $1p_{1/2}$ orbital, as shown the red solid curve in Fig.~\ref{fig:li10-angular-distributions-fig}. A $SF$ of 0.67 $\pm$ 0.12 is determined for the $(1p_{1/2})^{2}$ component. This result is much less than the prediction from the conventional shell model ($\simeq$ 2.0), confirming a relatively small $(p_{1/2})^{2}$ component in the ground state of ${\rm ^{11}{Li}}$. Assuming the remaining probability fraction to be $s$- and $d$-waves, a large $(2s_{1/2})^{2}$ probability fraction $\ge 44 \SI{}{\percent}$ is deduced for the ground state of ${\rm ^{11}{Li}}$. This data is also theoretically analyzed in Ref. \cite{bib:96} and a similar $p_{1/2}$-wave component, \SI{31}{\percent}, is given.

\begin{figure}[!htb]
\includegraphics
  [width=0.7\hsize]
  {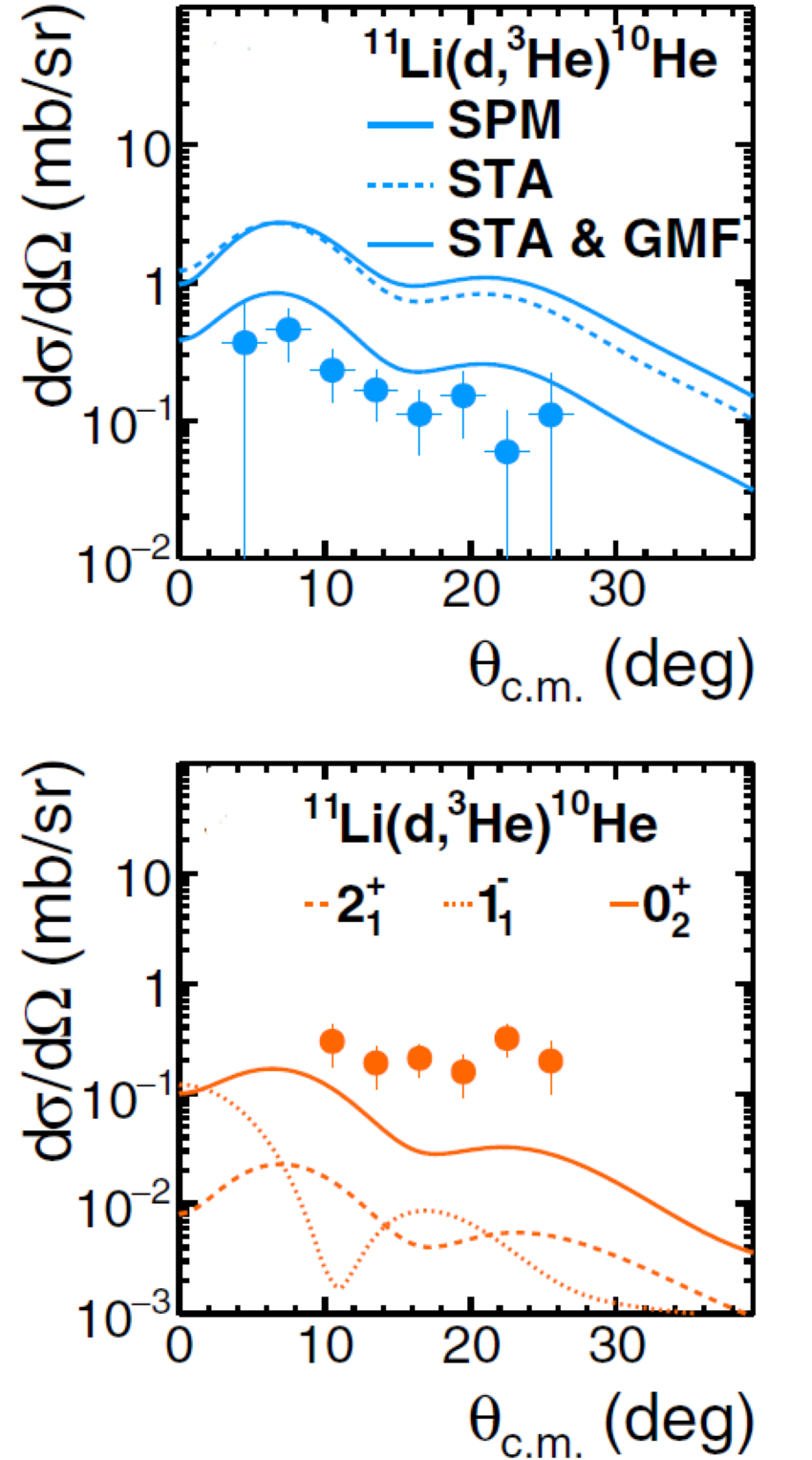}
\caption{Experimental ($d$, ${\rm ^{3}{He}}$) cross sections for populating the ground state (up) and excited state (down) of ${\rm ^{10}{He}}_{g.s.}$ in comparison with the DWBA calculations. SPM, STA and STA-GMF stand for different overlap functions used in the DWBA calculations. See more details in the text of Ref. \cite{bib:6}, which is also the source of this figure.}
\label{fig:he10-angular-distributions-fig}
\end{figure}

Recently, a $d$($^{11}$Li, ${\rm ^{3}{He}}$) reaction was performed at RIKEN in inverse kinematics with a radioactive beam of $^{11}$Li at 50 MeV/nucleon \cite{bib:6}. This reaction can be used to study the configurations in the ground state of ${\rm ^{11}{Li}}$.  The MM spectrum gated on $^8$He residuals shows two peaks at $1.4(3)$ and $6.3(7)$ \SI{}{\mega\electronvolt}. Angular distributions of ${\rm ^{3}{He}}$ nuclei in coincidence with ${\rm ^{8}{He}}$ residues, shown in Fig.~\ref{fig:he10-angular-distributions-fig}, were analyzed with the DWBA calculations. The OPs for the entrance channel of $^{11}$Li + $d$ were obtained from the elastic scattering data measured at the same experiment. The DWBA might not be an appropriate method to deal with the unbound ${\rm ^{10}{He}}$. Thus, it is argued that the removal of a deeply bound proton in the reaction justifies the simplification of the reaction model. The $<{\rm ^{11}{Li}}|{\rm ^{10}{He}}>$ overlap function used in the DWBA calculations was calculated from the standard potential model (SPM), source term approach (STA), and STA corrected by Geometrical
mismatch factor (GMF). The shape of the angular distribution is well reproduced, but the magnitude of the cross section for the ground state of ${\rm ^{10}{He}}$ is overestimated in all three cases. The cross sections for the excited states are larger than the DWBA predictions with the assumed $SF$s from shell model calculations for the $2^{+}_{1}$, $1^{-}_{1}$, $0^{+}_{0}$ states in ${\rm ^{10}{He}}$. Those experimental results may be attributed to the important contribution of ${\rm ^{10}{He}}$ core excitation in the ground-state wave function of ${\rm ^{11}{Li}}$.

The ground state of ${\rm ^{11}{Li}}$ was also studied by a 2$n$ transfer reaction, $p$(${\rm ^{11}{Li}}$, $t$)${\rm ^{9}{Li}}$ \cite{bib:72},  with a radioactive beam of  ${\rm ^{11}{Li}}$ at a very low incident energy of 3 MeV/nucleon. The experiment was performed \cite{bib:72} at TRIUMF by using the active target detector MAYA. Multi-step transfer calculations were applied with different wave functions of ${\rm ^{11}{Li}}$. Only the wave functions with strong mixing of $p$ and $s$ neutrons  as well as the three-body correlations, can provide the best fit to the magnitude of the DCSs.

\subsection{Beryllium isotopes}

\textbf{$^{11}$Be}

Since the beta decay experiment of ${\rm ^{11}{Be}}$ \cite{bib:76} indicating that the spin-parity of the ground state of ${\rm ^{11}{Be}}$ is ${{1/2}^{+}}$ \cite{bib:77} instead of ${{1/2}^{-}}$, which means that the last neutron prefers to take the $2s_{1/2}$ orbital instead of the $1p_{1/2}$ orbital, the vanishing of the $N$ = 8 shell closure has been proposed to explain such phenomenon in neutron-rich light nuclei. A spectroscopic study via $d$(${\rm ^{10}{Be}}$, $p$) reaction performed at Oak Ridge National Laboratory with a radioactive beam of ${\rm ^{10}{Be}}$ at four different incident energies, shows that the average $SF$ for a neutron in an ${nlj = 2s_{1/2}}$ state coupled to the ground state of ${\rm ^{10}{Be}}$ is 0.71 $\pm$ 0.05 \cite{Schmitt}. This large $SF$ indicates a large $s$-wave strength in the ground-state of ${\rm ^{11}{Be}}$.

In addition to the naive ${n \otimes {\rm ^{10}Be(0^{+}_{gs})}}$ component, the ground state wave function of ${\rm ^{11}{Be}}$ was found to have a considerable overlap with a valence neutron coupled to an excited $\rm ^{10}Be$ $(2^{+})$ core via a ($p, d$) transfer reaction with a ${\rm ^{11}{Be}}$ beam \cite{Winfield, bib:79, bib:80}. For the former configuration, the valence neutron surrounding the inert core ${\rm ^{10}{Be}}$ populates the intruder $s$-orbital other than the normal $p$-orbital. However, in the latter case, ${\rm ^{10}{Be}}$ is excited to the ${2^{+}}$ state at \SI{3.37} {\mega\electronvolt} and the valence neutron fills into the intruder $d$-orbital. The $s$- and $d$-wave components in the ground state of ${\rm ^{11}{Be}}$ were deduced to be \SI{84}{\percent} and \SI{16}{\percent}, respectively, from the ($p$, $d$) transfer reaction with the beam energy at \SI{35.3}{\mega\electronvolt}/nucleon \cite{Winfield, bib:79, bib:80}.  The results were confirmed by the same reaction with a radioactive ${\rm ^{11}{Be}}$ beam at \SI{26.9}{\mega\electronvolt}/nucleon \cite{Jiang-2018}.

\textbf{$^{12}$Be}

With one more neutron, the valence-nucleon configuration of ${\rm ^{12}{Be}}$ also attracts a lot of people's attention. Different from $^{11}$Li, besides $s$- and $p$-orbital, two valence neutrons may populate $d$-orbital. But, which component is dominant? This question has been studied through several $d$(${\rm ^{11}{Be}}$, $p$) reactions.

In the $d$(${\rm ^{11}{Be}}$, $p$) reaction performed at TRIUMF \cite{Kanungo}, the $s$-wave neutron fraction of the first two $0^{+}$ levels in ${\rm ^{12}{Be}}$ was investigated for the first time. In this experiment, three peaks were observed in the excitation energy spectrum, including the ground state and a $1^{-}$ states located at \SI{2.71}{\mega\electronvolt}, as well as a unresolved doublet made up of the $0_{2}^{+}$ and $2^{+}$ states. Angular distributions for each peak in excitation energy spectrum were analyzed using the DWBA method with global optical model potentials as well as parameters from the neighboring nuclei. The ground state $s$-wave $SF$ was determined to be $0.28^{+0.03}_{-0.07}$ while that for the long-lived $0_{2}^{+}$ excited state was $0.73^{+0.27}_{-0.40}$. The value for the $0_{2}^{+}$ state was given with a large uncertainty, because this state was not clearly distinguished from the $2^{+}$ state. The result has been questioned by H. T. Fortune \emph{et al} \cite{bib:83}, as it is inconsistent with the results from knock-out experiments and theoretical calculations.

The same reaction was studied with a lower beam energy (2.8 MeV/nucleon) at the REX-ISOLDE facility \cite{Johansen}. Besides the outgoing proton measured by the T-REX silicon detector array, the $\gamma$-rays emitting from the excited states in ${\rm ^{12}{Be}}$ were also detected by the MINIBALL germanium array. The $\gamma$-ray detection enabled a clear identification of the four known bound states in ${\rm ^{12}{Be}}$, and the angular distribution for each state has been studied individually. In this case the $SF$ for the ground state is also smaller than for the excited $0_{2}^{+}$ state. This experiment suffered from a very low beam energy, leading to an ineffective detection inside the most sensitive angular range, especially for the $0_{2}^{+}$ state.

\begin{figure}[!htb]
\includegraphics
  [width=0.8\hsize]
  {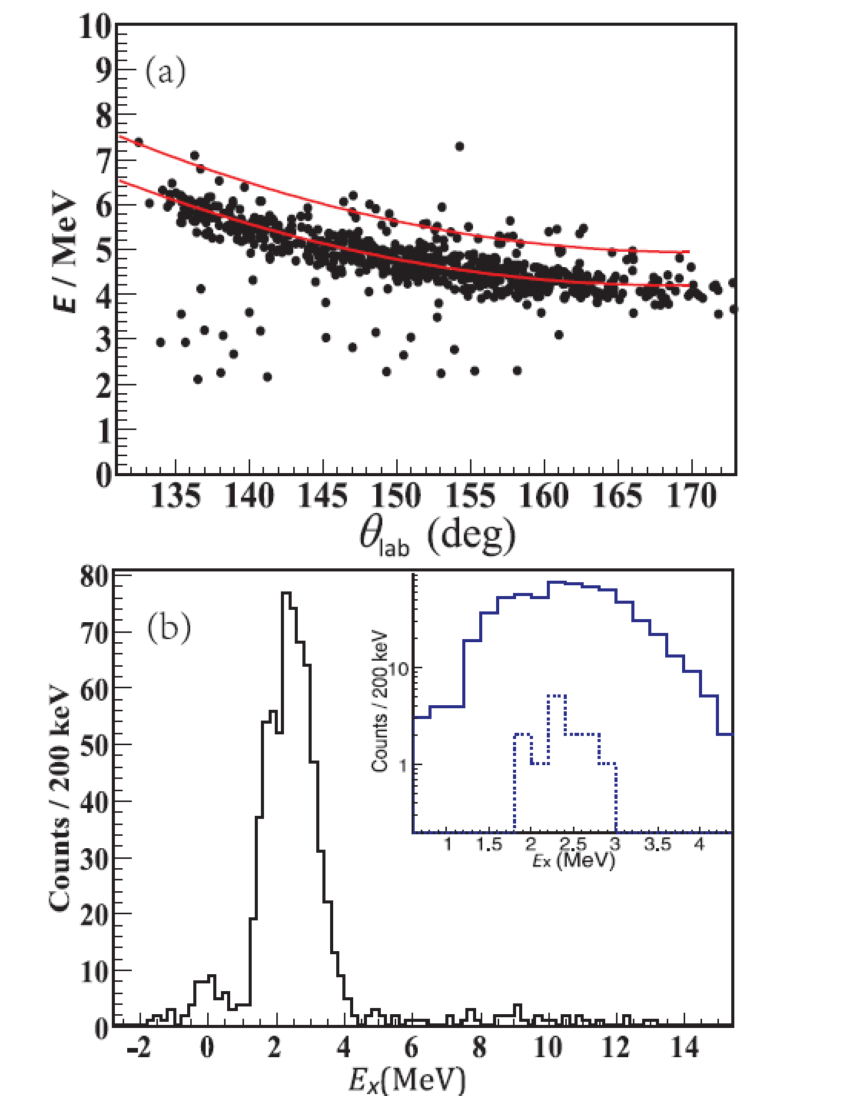}
\caption{(a) The measured proton energies versus the laboratory angles, gated on the ${\rm ^{12}{Be}}$ in the zero degree silicon detector. The red solid lines illustrate the calculated kinematics of the $d$(${\rm ^{11}{Be}}$, $p$) transfer reaction to the ground state and the \SI{2.251}{\mega\electronvolt} excited state. (b) The excitation energy spectrum of ${\rm ^{12}{Be}}$ deduced from the recoil protons in (a). The dotted curve in the inset shows the events in coincidence with the \SI{0.511}{\mega\electronvolt} $\gamma$ rays detected by the scintillation counters. This figure is from Ref. \cite{Chen4}.}
\label{fig:be12-rcnp-energy-fig}
\end{figure}

\begin{figure}[!htb]
\includegraphics
  [width=0.95\hsize]
  {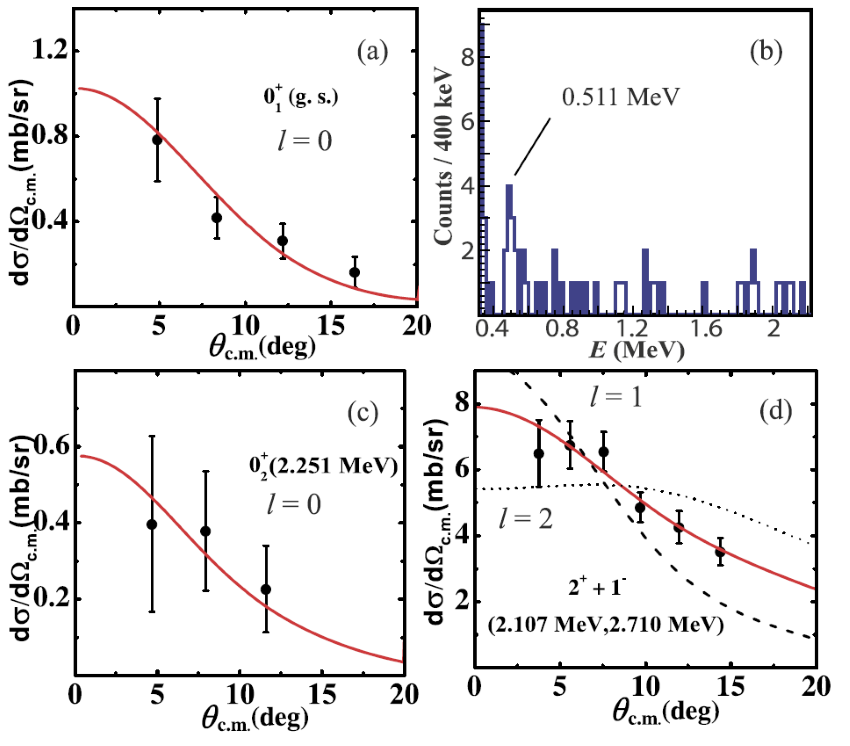}
\caption{Experimental DCSs of the ${\rm ^{11}{Be}}$($d$, $p$) reaction with a radioactive beam of $^{11}$Be at 26.9 MeV/u  (solid dots), together with the FR-ADWA calculations (curves), for (a) the $g.s.$ ($0^{+}_{1}$), (c) the isomeric state ($0^{+}_{2}$), and (d) the summed $2^{+}$ and $1^{-}$ states. $l$ in (a), (c) and (d) denotes the transferred orbital angular momentum into the final state of ${\rm ^{12}{Be}}$. (b) is dedicated to the $\gamma$-ray energy spectrum in coincidence with ${\rm ^{12}{Be}}$ + $p$ events. This figure is from Ref. \cite{Chen-PLB}.}
\label{fig:be12-rcnp-angular-distributions-fig}
\end{figure}

Due to the lack of proper normalization procedures for the two transfer reactions stated above, it would be difficult to compare their $SF$ results with other measurements or to each other \cite{Kay-2013}. With the purpose to further study the intruder configuration in ${\rm ^{12}{Be}}$, a new measurement of the $d$(${\rm ^{11}{Be}}$, $p$) reaction was performed at the EN-course beam line, RCNP, with special measures taken to deal with the questioned experimental uncertainties of the two previous experiments \cite{Chen-PLB}. Fig.~\ref{fig:rcnp-12be-setup-fig} shows the schematic view of the experimental setup. A special isomer-tagging method was used to discriminate the $0_{2}^{+}$ state from the broad excitation-energy peak. Fig.~\ref{fig:be12-rcnp-energy-fig} (a) shows the measured proton energies versus the laboratory angles, gated on the ${\rm ^{12}{Be}}$ in the zero degree silicon detector. Fig.~\ref{fig:be12-rcnp-energy-fig} (b) gives the excitation energy spectrum of ${\rm ^{12}{Be}}$ deduced from the energies and angles of the recoil protons. Elastic scattering of ${\rm ^{11}{Be}}$ + $p$ was simultaneously measured to estimate the hydrogen contamination in the ${\rm (CD_{2})}_{n}$ target and to obtain the reliable OP to be used in the analysis of the transfer reaction. The FR-ADWA calculations were employed to extract the $SF$s for the low-lying states in ${\rm ^{12}{Be}}$. Fig.~\ref{fig:be12-rcnp-angular-distributions-fig} shows the comparison of experimental and theoretical cross sections. The extracted $s$-wave $SF$s are $0.20^{+0.03}_{-0.04}$ and $0.41^{+0.11}_{-0.11}$ for the $0^{+}_{1}$ and $0^{+}_{2}$ states, respectively. The ratio between the $SF$s of the first two low-lying $0^{+}$ states, together with the previously reported results for the $p$-wave components, were used to deduce the single-particle intensities in these two bound $0^{+}$ states of ${\rm ^{12}{Be}}$, which can be compared directly with shell model predictions.  The $s$-, $d$-, $p$-wave intensity for the $0^{+}_{1}$ and $0^{+}_{2}$ states are deduce to be 0.19 $\pm$ 0.07, 0.57 $\pm$ 0.07, 0.24 $\pm$ 0.05, and 0.39 $\pm$ 0.02, 0.02 $\pm$ 0.02, 0.59 $\pm$ 0.05, respectively. The error bars are deduced from the statistic uncertainties of the $SF$s. The results show a small $s$-wave (0.19 $\pm$ 0.07) but a obviously dominant $d$-wave(0.57 $\pm$ 0.07) intruding in the ground state of ${\rm ^{12}{Be}}$, which is dramatically different from the ground state of ${\rm ^{11}{Be}}$ predominated by an intruder $s$-wave. The experimental results are compatible with those obtained from the previous transfer reaction measurements, considering the reported uncertainties.

In principal, there should be three $0^{+}$ states in $^{12}$Be in this $p-sd$ model space. Until now, only the lowest two have been found in the bound region, but the third $0^{+}$ has not been identified experimentally. More studies, such as $^{13}$B($d$, $^3$He)$^{12}$Be or $^{14}$B($d$, $^4$He)$^{12}$Be, are encouraged to further investigate the excited states in ${\rm ^{12}{Be}}$.

\subsection{Boron isotopes}

For the $N$ = 8 system, as reviewed above, the halo nucleus $^{11}$Li has a large $s$-wave component in
its ground-state wave function. In contrast, the larger $s$-wave component was observed in the 0$_2^+$ isomeric state rather than the 0$_1^+$ ground state in $^{12}$Be. How about the nucleus $^{13}$B? Its ground state
has $J^{\pi}$ = $3/2^-$, and the properties of its low-lying positive parity
states contain a number of information about the $2s_{1/2}$ and $1d_5/2$
single-particle energies and the residual interaction.
For the $N$ = 9 isotones, $^{12}$Li and $^{13}$Be are particle unbound, but $^{14}$B with $S_n$ = 0.97 MeV is loosely bound. Therefore, the lightest particle-bound $N$ = 9 isotone, $^{14}$B, provides an unique opportunity to study the evolution of the properties of single-neutron states, especially the gap between the $2s_{1/2}$ and $1d_5/2$ orbitals.
The $N$ = 10 isotones are ideal for studying two-neutron interactions in the $sd$ shell. From the knock-out experiment, it was found that the ground state of $^{14}$Be is predominated by $s$-wave (87$\%$) \cite{bib:69}. From the $^{15}$C($d$, $p$) transfer reaction, the 0$_1^+$ ground state of $^{16}$C was found to be dominated by $d$-wave \cite{16C-Wuosmaa}. How about $^{15}$B, which is between $^{14}$Be and $^{16}$C? What's the percent of $s$-wave component in the ground state wave function of $^{13,14,15}$B? Does $s$-wave dominates in their low-lying states? In order to answer these questions, several transfer reaction experiments were performed.

\textbf{$^{12,13}$B}

\begin{figure}[!htb]
\includegraphics
  [width=0.8\hsize]
 {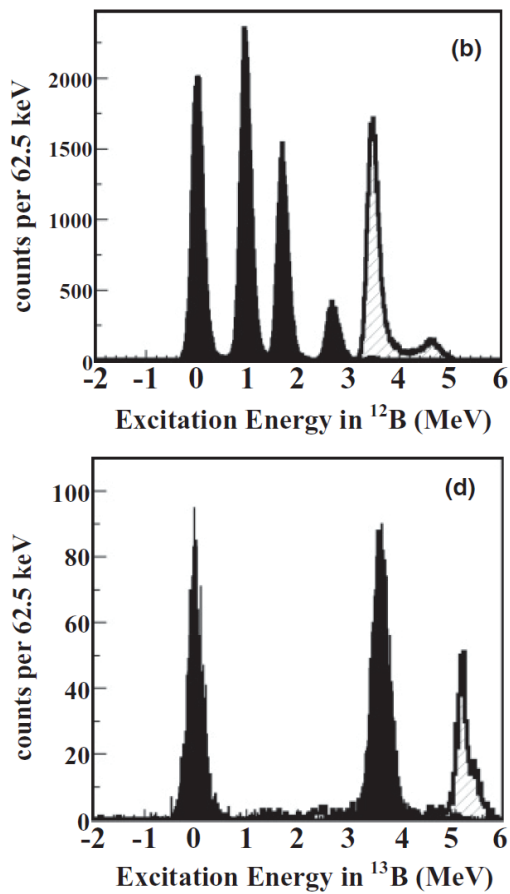}
\caption{The excitation energy spectra for $^{12}$B (upper) and $^{13}$B (lower) from the ${d}$(${\rm ^{11}{B}}$, ${p}$) and ${d}$(${\rm ^{12}{B}}$, ${p}$) transfer reactions, respectively \cite{bib:73}. The solid and dashed shadows stand for the bound and unbound excited states, which were constructed from the recoil protons in coincident with $^{12}$B ($^{13}$B) and $^{11}$B ($^{12}$B) in the upper (lower) figure, respectively. }
\label{12b13bexcitation}
\end{figure}

The experiment of ${d}$(${\rm ^{11,12}{B}}$, ${p}$) transfer reaction to the low-lying states of ${\rm ^{12,13}{B}}$ was performed at Argonne with silicon detector \cite{bib:73}. In Fig. \ref{12b13bexcitation}, the excitation energy spectra for $^{12}$B (upper) and $^{13}$B (lower), which were constructed by using the recoil protons in coincidence with $^{12}$B and $^{13}$B, are shown. The overall $Q$-value resolution is
approximately 250 keV (FWHM), which is not good enough to discriminate the doublet state at  $E_{x}=3.482$ and \SI{3.681}{\mega\electronvolt} in ${\rm ^{13}{B}}$. This doublet is expected to contain the  $s$- and $d$-wave components.

\begin{figure}[!htb]
\includegraphics
  [width=0.95\hsize]
  {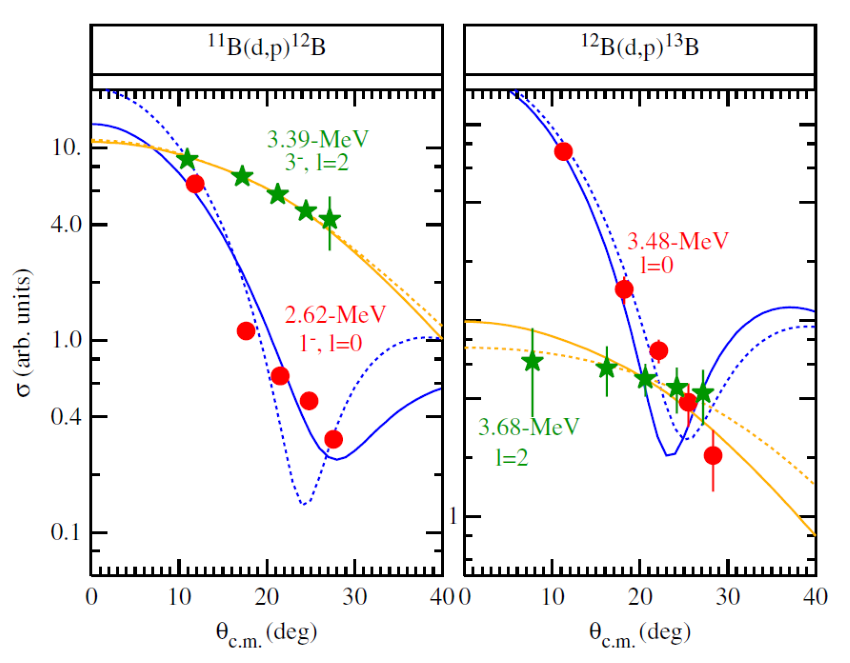}
\caption{The left-hand and right-hand figures show angular distributions for the ${\rm ^{11}{B}}$($d$, $p$)${\rm ^{12}{B}}$ and ${\rm ^{12}{B}}$($d$, $p$)${\rm ^{13}{B}}$ reaction, respectively, for two states known to be populated with $l = 0$ and $l = 2$. The solid and the dashed lines are the DWBA calculations with two different sets of parameters \cite{bib:28}. }
\label{fig:boron13-angular-distributions-fig}
\end{figure}

A new experiment of ${d}$(${\rm ^{11,12}{B}}$, ${p}$) transfer reaction to the low-lying states of ${\rm ^{12,13}{B}}$ was performed at Argonne with the HELIOS setup  \cite{bib:28}.  The properties of some low-lying states in the neutron-rich $N$ = 8 nucleus ${\rm ^{13}{B}}$ were studied \cite{bib:28}. Two closely spaced states at $E_{x}=3.482$ and \SI{3.681}{\mega\electronvolt} in ${\rm ^{13}{B}}$ were observed and clearly identified from each other, which benefits from the better energy resolution of HELIOS. The angular distributions are shown in Fig.~\ref{fig:boron13-angular-distributions-fig}. The state at \SI{3.48}{\mega\electronvolt} shows $l = 0$ angular distribution and the spin-parity is assigned as $(1/2)^+$. The angular distribution of the state located at \SI{3.68}{\mega\electronvolt} excitation energy, exhibits a $l = 2$ shape. Based on a comparison with shell model calculations, $J^\pi = (5/2)^+$ is tentatively assigned.

\begin{figure}[!htb]
\includegraphics
  [width=0.90\hsize]
  {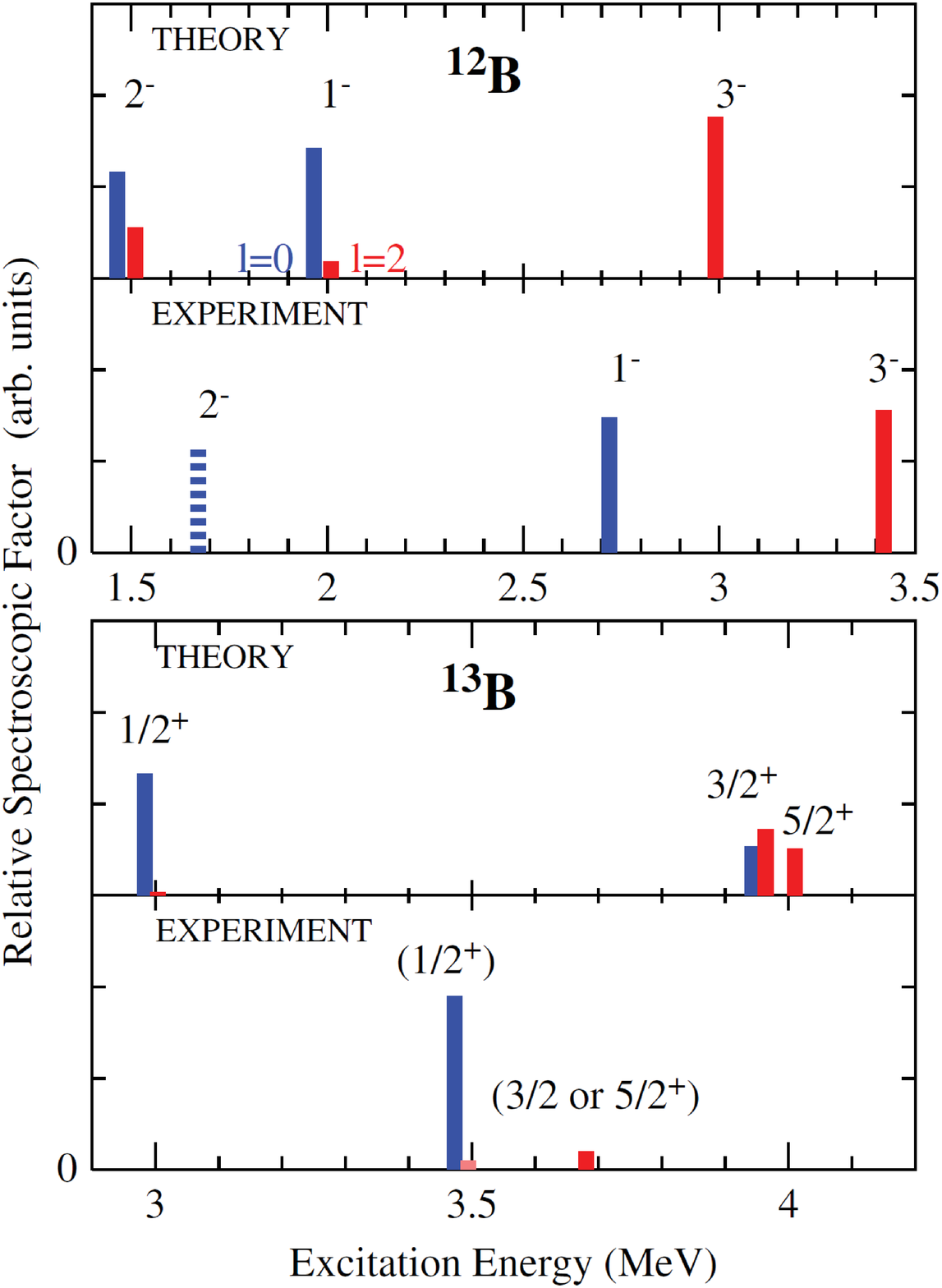}
\caption{Comparing with the shell-model calculations, the relative $SF$ for $l = 0$ (blue bars) and $l = 2$ (red bars) of the $d$(${\rm ^{11}{B}}$ , $p$)${\rm ^{12}{B}}$ (upper) and $d$(${\rm ^{12}{B}}$, $p$)${\rm ^{13}{B}}$ (lower) reactions. Solid bars are experimental values
from the measurement in Ref. \cite{bib:28}, while the dashed bar is from the
work published in Ref. \cite{bib:73} . }
\label{13Bstrength}
\end{figure}

Absolute normalization of the cross sections was not possible, because the absolute beam intensity was not measured in this experiment. Only the relative $SF$s are meaningful. As shown in Fig. \ref{13Bstrength}, the relative $SF$ of the suggested $(5/2)^+$ state, compared with that for the $(1/2)^+$ state, is smaller than the value predicted by shell model. More $d$-wave strength is expected at the excited states above \SI{4}{\mega\electronvolt} in ${\rm ^{13}{B}}$. Seen from Fig. \ref{13Bstrength}, the absolute excitation energies, as well as the ordering of the excited states, are not in good agreement with shell-model predictions. A predicted $(3/2)^+$ state is not observed below \SI{4.8}{\mega\electronvolt}, which maybe correspond to a strong transition at \SI{5.1}{\mega\electronvolt} seen in a prior measurement (the lower picture in Fig. \ref{12b13bexcitation}) \cite{bib:73}. More positive parity states are expected to exist above the neutron separation energy and further experimental studies are encouraged.

\begin{figure}[!htb]
\includegraphics
  [width=0.80\hsize]
  {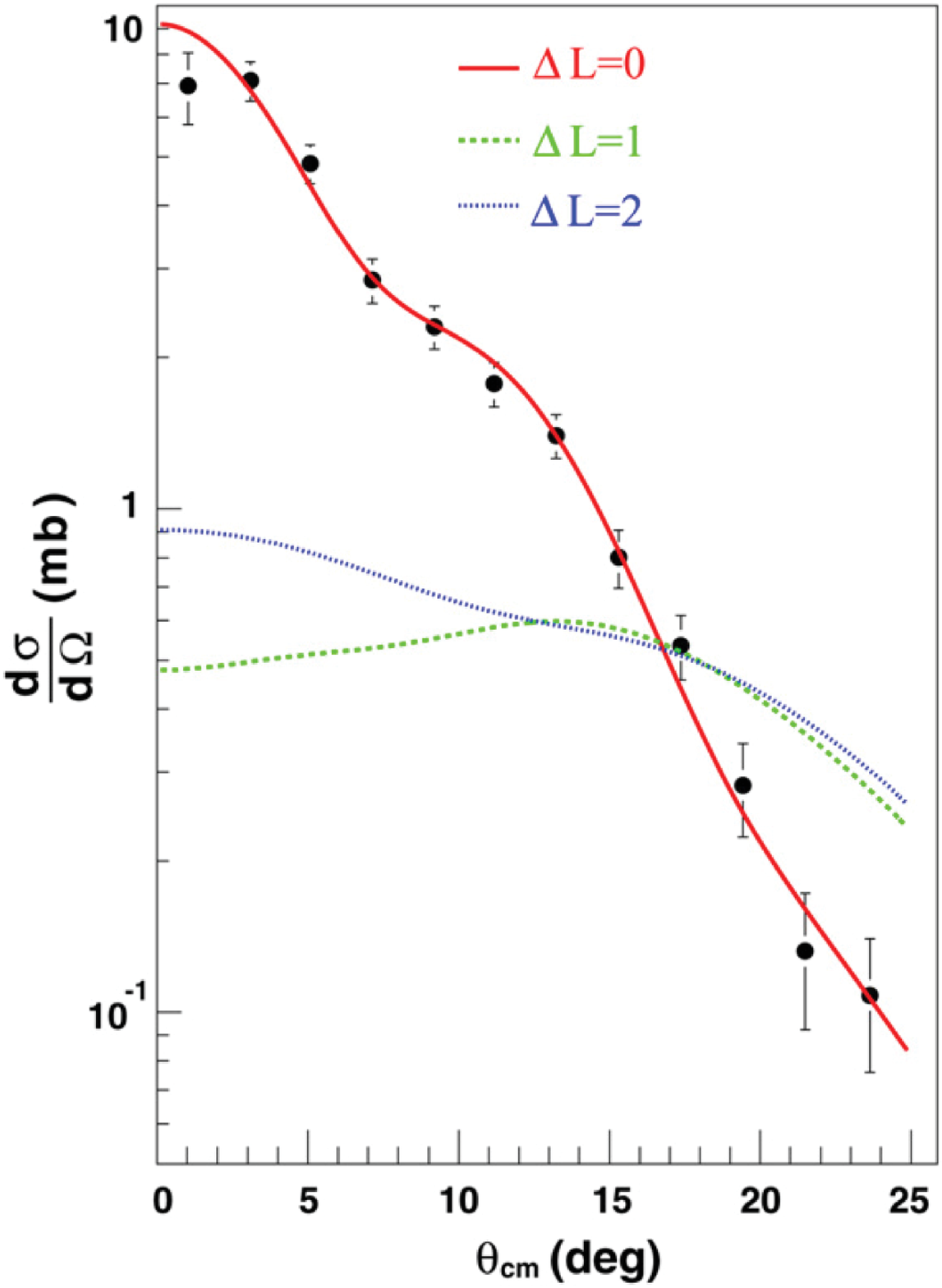}
\caption{Angular distributions for the  $^4$He($^{12}$Be, $ ^{13}$B$\gamma$) reaction to the $E_x$ = 4.83 MeV resonant state. The optical model potentials for the entrance and exit channels are obtained from the folding model \cite{bibxx1}. }
\label{fig:12be4he13b}
\end{figure}

A state at $E_x$ = 4.83 MeV is strongly populated in the $^4$He($^{12}$Be, $ ^{13}$B$\gamma$) reaction with a secondary beam of $^{12}$Be at 50 MeV/nucleon \cite{bibxx1}.  The spin and parity of this state were assigned to 1/2$^+$ by comparing the DCSs data with the DWBA calculations, see Fig. \ref{fig:12be4he13b}. The DWBA predictions with $\Delta l$ = 0, 1, 2 are shown as red solid, green dashed and blue dotted curves in Fig. \ref{fig:12be4he13b}. The angular distributions, especially the peak at forward angles, are well
described by the $\Delta l$ = 0 DWBA calculations. Thus, it is interpreted as a proton intruder state. In $^{13}$B, another 2$p$-2$h$ state at $E_x$ = 3.53 MeV with a tentatively assigned spin-parity of $3/2^-$ was suggested to be a neutron intruder state \cite{13bneutronintruder}.

\textbf{$^{14}$B}

\begin{figure}[!htb]
\includegraphics
  [width=0.8\hsize]
  {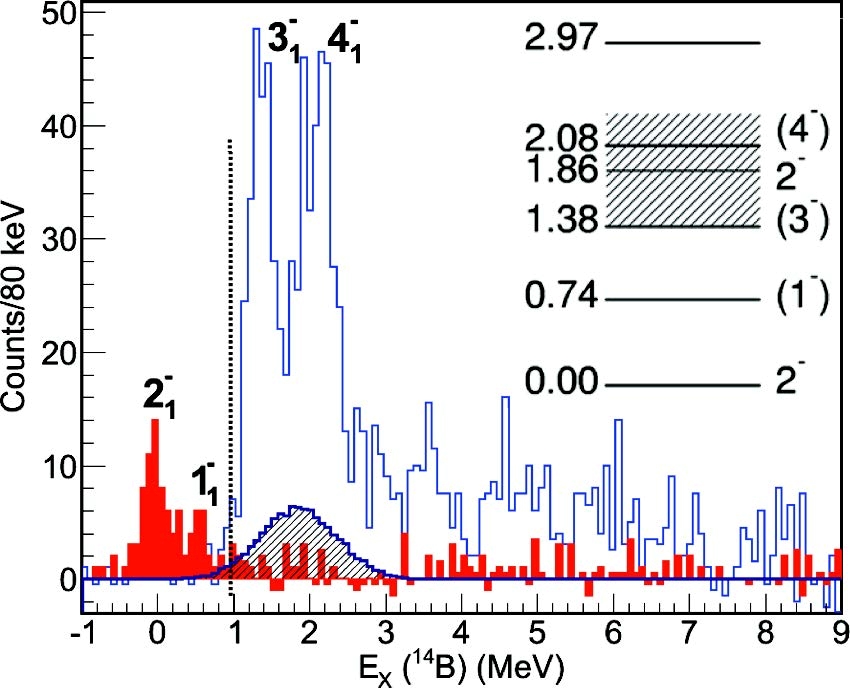}
\caption{${\rm ^{14}{B}}$ excitation-energy spectrum from the $d$(${\rm ^{13}{B}}$, $p$)${\rm ^{14}{B}}$ reaction. The filled (open) histogram corresponds to protons detected in coincidence with identified ${\rm ^{14}{B}}$(${\rm ^{13}{B}}$) recoil ions. The vertical dashed line shows the neutron-separation energy. The inset shows the level diagram for ${\rm ^{14}{B}}$ \cite{Bedoor}.}
\label{fig:boron14-energy-fig}
\end{figure}

The experimental data of $d$(${\rm ^{13}{B}}$, $p$)${\rm ^{14}{B}}$ reaction using HELIOS, confirm that the ground and first-excited states are predominantly populated by $s$-wave in character and are single-neutron halo states \cite{Bedoor}.
 The normalized $SF$s for the low-lying states are shown in Tab. \ref{13bdp}, suggesting that the  $2^{-}$ ground state has a moderate $2s_{1/2} - 1d_{5/2}$ configuration mixing and the $1^{-}$ first excited state is nearly pure $2s_{1/2}$. The effective single-particle energies of the $2s_{1/2}$ and $1d_{5/2}$ neutron orbitals in $^{14}$B, show the inversion of these two orbitals compared to the normal ordering in the valley of stability \cite{Bedoor}. Fig.~\ref{fig:boron14-energy-fig} shows the excitation-energy spectrum of ${\rm ^{14}{B}}$ obtained in this experiment. The $2_2^-$ and $1_2^-$ states, which are expected to be comprised of one $1p_{3/2}$ valence proton (hole) coupling with one $1d_{5/2}$ valence neutron, were not observed. However, the unbound $3_1^-$ and $4_1^-$ states, made up of the same coupling of $\pi(1p_{3/2})^{-1}\nu (1d_{5/2})^1$, were obviously measured.

\begin{figure}[!htb]
\includegraphics
  [width=0.8\hsize]
  {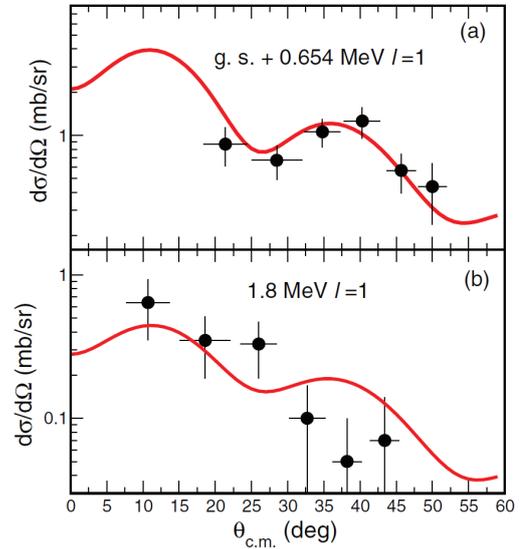}
\caption{Angular distribution for the $^{15}$C($d$, ${\rm ^{3}{He}}$) reaction to $2_1^--1_1^-$ doublet (a) and the $2_2^-$ state at $E_x$ = 1.8 MeV (b) in $^{14}$B \cite{bib:68}.}
\label{fig:14bangulardistribution3he}
\end{figure}

Furthermore, with the same device, HELIOS, the ($d$, ${\rm ^{3}{He}}$) experiments were performed with ${\rm ^{14}{C}}$ and ${\rm ^{15}{C}}$ secondary beams, providing spectroscopic information about the final states in ${\rm ^{13}{B}}$ and ${\rm ^{14}{B}}$ \cite{bib:68}. For the ${\rm ^{14}{C}}$ data,
  several transitions, which were reproduced well with the transferred angular momentum $l$ = 1, are observed.  This result indicates that the ground state of ${\rm ^{14}{C}}$ likely possesses $1p_{1/2}$ proton character. A very weak $l$ = 0 transition is also measured, which may be associated with a possible ${1/2}^{+}$ proton intruder state in ${\rm ^{13}{B}}$ \cite{bibxx1}. It is worth noting that such an excitation is not well described by shell-model calculations. For the ${\rm ^{15}{C}}$ data, the ground state ($2_1^-$) and the first excited state ($1_1^-$) at $E_x$ = 0.654 MeV in $^{14}$B were not discriminated from each other due to a worse energy resolution of HELIOS for $^3$He. In addition to the bound states, a broad unbound excited $2_2^{-}$ state at around $E_x$ = 1.8 MeV in $^{14}$B, which was not seen but was expected in the previous $d$(${\rm ^{13}{B}}$, $p$)${\rm ^{14}{B}}$ experiment \cite{Bedoor}, was observed clearly by the $d$($^{15}$C, ${\rm ^{3}{He}}$) reaction. Fig.~\ref{fig:14bangulardistribution3he} shows the angular distributions of $^{15}$C($d$, ${\rm ^{3}{He}}$) to the bound $2_1^--1_1^-\mathcal{}$ doublet and the unbound $2_2^-$ state in $^{14}$B. All the DCSs are reasonably well described by $l$ = 1 proton removal from $^{15}$C.
    The relative $SF$s extracted from this experiment are consistent with shell-model calculations, as well as expectations based on prior measurements of the $d$(${\rm ^{13}{B}}$, $p$)${\rm ^{14}{B}}$ reaction.

\textbf{$^{15}$B}

Previous experiments have shown a large $s$-wave component in the ground state of ${\rm ^{14}{B}}$. With one more neutron, it is an interesting question about how much $s$-wave component in the ground state and low-lying excited states of its adjacent isotope ${\rm ^{15}{B}}$, but there are no experimental data to answer this question till now. Recently, ($d$, $^3$He) and ($d$, $p$) transfer reactions in inverse kinematics were performed to study the spectroscopic information of ${\rm ^{15}{B}}$. In principle, there should be two $J^{\pi}=(3/2)^{-}$ states, including the ground state, in ${\rm ^{15}{B}}$ due to the mixture of $s$- and $d$-wave components. But the second $(3/2)^{-}$ has not been found experimentally. The $s$-wave $SF$s for the known states, including the ground state of ${\rm ^{15}{B}}$ are unknown. In theory, shell model predicted a small $s$-wave $SF$ of 0.48 for the ground state \cite{Tanihata}, dramatically smaller than the expected value 2.0, which implies that there is a larger $s$-wave $SF$ existing in the excited $(3/2)^{-}$ state, and seems to be same as the situation of ${\rm ^{12}{Be}}$ and ${\rm ^{16}{C}}$.

\begin{figure}[!htb]
\includegraphics
  [width=0.8\hsize]
  {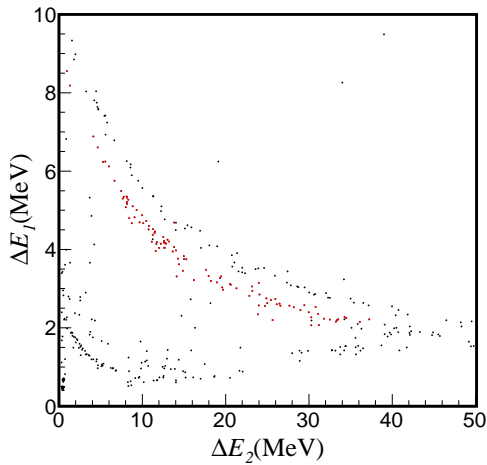}
\caption{Particle identification of the light particle in coincidence with the ${\rm ^{15}{B}}$ detected at zero degree. $^3$He particles are shown as red dots.}
\label{fig:c16-he3-2018-fig}
\end{figure}

The $d$(${\rm ^{16}{C}}$, $^3$He) ${\rm ^{15}{B}}$ experiment was recently finished at RIBLL1 in Lanzhou, China. The preliminary result of the light particle PID is shown in Fig.~\ref{fig:c16-he3-2018-fig}. With coincidence of $^{15}$B, some $^3$He particles recoil from this transfer reaction are clearly seen. Further data analysis is in progress.

\begin{figure}[!htb]
\includegraphics
  [width=0.8\hsize]
  {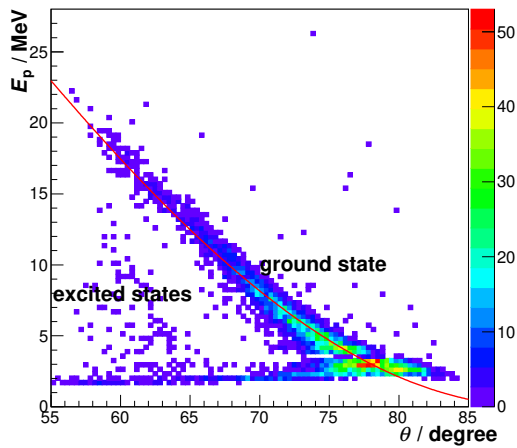}
\caption{Energy versus angle for protons scattered elastically from $^{14}$B. The red line shows the calculated kinematic curve. }
\label{fig:b14-p-2018-fig}
\end{figure}

\begin{figure}[!htb]
\includegraphics
  [width=0.8\hsize]
  {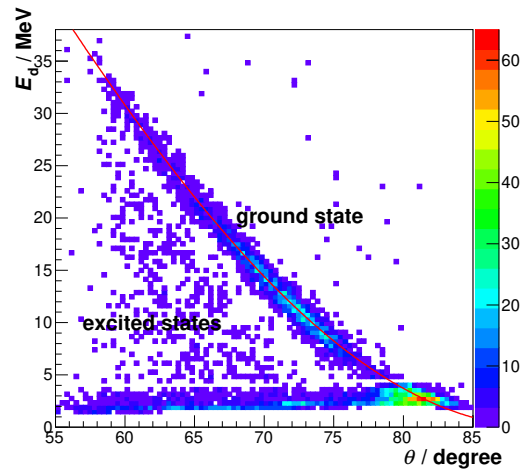}
\caption{Energy versus angle for elastic scattering of $^{14}$B + $d$. The red line shows the calculated kinematic curve for this reaction.}
\label{fig:b14-d-2018-fig}
\end{figure}

\begin{figure}[!htb]
\includegraphics
  [width=0.8\hsize]
  {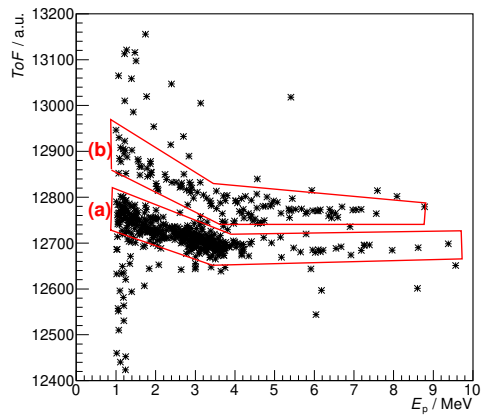}
\caption{Particle identification of annular silicon detector using the TOF-$\Delta E$ method. (a) Protons from transfer reaction. (b) Protons from the fusion-evaporation reaction between the secondary beam $^{14}$B and the silicon detectors placed at around zero degree relative to the beam line.}
\label{fig:b14-dp-p-2018-fig}
\end{figure}

Shortly after the ($d$, $^3$He) experiment, the $d$(${\rm ^{14}{B}}$, $p$) ${\rm ^{15}{B}}$ experiment was performed at RCNP, which used a nearly similar setup as shown in Fig.~\ref{fig:rcnp-12be-setup-fig}. The elastic scattering of $^{14}$B on the proton and deuteron targets were measured at the same experiment in order to extract OPs for the theoretical calculations of transfer reaction. The energy versus angle for protons and deuterons emitting from the elastic scattering channels are shown in Fig.~\ref{fig:b14-p-2018-fig} and Fig.~\ref{fig:b14-d-2018-fig}, respectively. The protons from transfer reaction were measured by a single-layer annular silicon detector at the backward angles, and the PID is achieved by using the TOF-$\Delta E$ method. As shown in Fig.~\ref{fig:b14-dp-p-2018-fig}, the proton from transfer reaction is clearly distinguished from other particles coming from the fusion-evaporation reaction between $^{14}$B and silicon detectors in the zero degree telescope.

\subsection{Carbon isotopes}
The intruder components have been studied for the neutron-rich Li, Be, B isotopes via transfer reactions. Now, let us review experimental studies on carbon isotopes. With 8 neutrons filling the $p$ shell and a high excitation energy (\SI{6.09}{\mega\electronvolt}) of the first excited ($2^{+}$) state, ${\rm ^{14}{C}}$ shows a good magic character for $N$ = 8. However, the spin-parity of ${\rm ^{15}{C}}$ ground state is $(1/2)^{+}$, which shows clearly that $2s_{1/2}$ orbital is below the $1d_{5/2}$ orbital. Studies of the ${\rm ^{14}{C}}$($d$, $p$)${\rm ^{15}{C}}$ reaction \cite{bib:75} show that the ground state of ${\rm ^{15}{C}}$ is well described as a valence neutron in a single-particle state around ${\rm ^{14}{C}}$ core, with the $s$-wave $SF$ as large as \SI{0.88}{}. It is also indicated that the ground state of ${\rm ^{15}{C}}$ has a small fraction of the ${\rm ^{14}{C}(2^{+})}\times{d}$ configuration \cite{bib:86}. This component is similar to the core excitation component in the ground state of $^{11}$Be.

\begin{figure}[!htb]
\includegraphics
  [width=0.9\hsize]
  {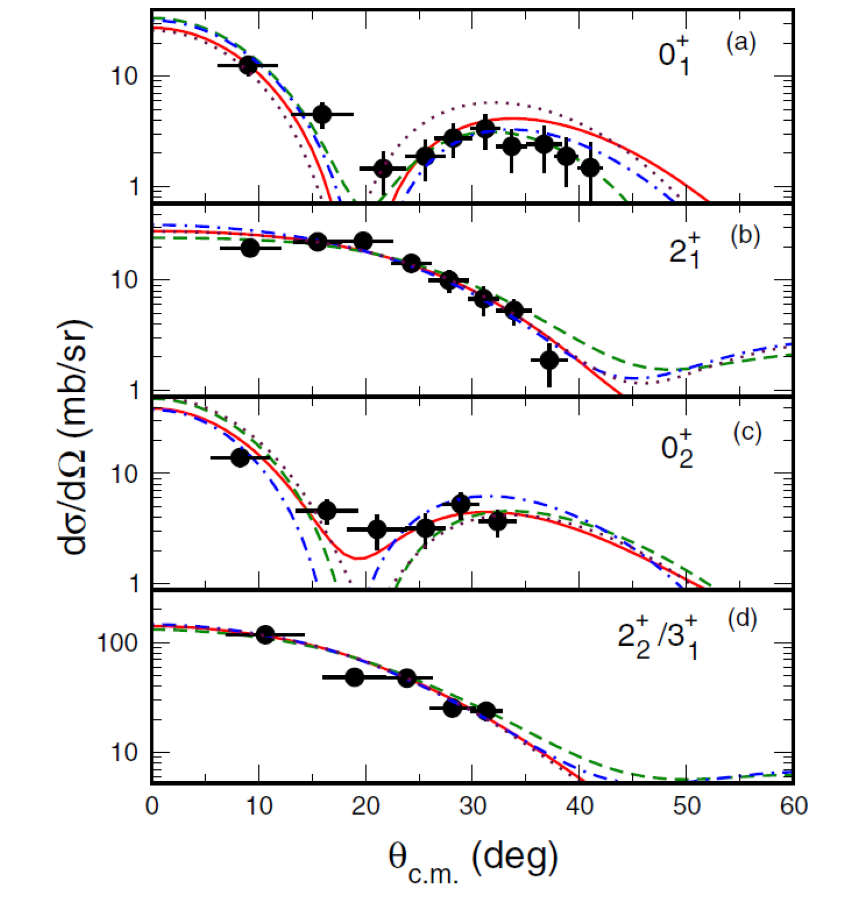}
\caption{Angular distributions for the ${\rm ^{15}{C}}$($d$, $p$) reaction to the $0_1^+$ (a), $2_1^+$ (b), $0_2^+$ (c) states and a $2_2^+/3_1^-$ doublet (d) in $^{16}$C. The curves represent DWBA calculations with different OP parameters \cite{16C-Wuosmaa}.}
\label{fig:carbon16-angular-distributions-fig}
\end{figure}

However, the study of $d$(${\rm ^{15}{C}}$, $p$)${\rm ^{16}{C}}$ reaction \cite{16C-Wuosmaa} shows a different picture, suggesting that ${\rm ^{16}{C}}$ may not need to be described with very exotic phenomena. The $d$(${\rm ^{15}{C}}$, $p$)${\rm ^{16}{C}}$ reaction was carried out at Argonne National Laboratory. A very thin deuterated polyethylene ((C$_2$D$_4$)$_n$) is bombarded by the ${\rm ^{15}{C}}$ beam with a high beam intensity of 1$\sim$2 $\times 10^6$ pps. The recoil protons were detected by the HELIOS spectrometer. The $0_1^+$, $2_1^+$, $0_2^+$ states and a $2_2^+/3_1^-$ doublet in $^{16}$C were observed in the excitation-energy spectrum reconstructed by the recoil protons. Although the resolution is approximately 140 keV (FWHM), it was still
insufficient to resolve the closely spaced $2_2^+/3_1^-$ doublet near $E_x$ = 4 MeV. Comparing with DWBA calculations using four sets of optical-model parameters, the angular distributions for these four populated states are shown in Fig. \ref{fig:carbon16-angular-distributions-fig}. Absolute $SF$s were obtained by comparing the experimental cross sections with DWBA calculations. The absolute $SF$s were normalized by requiring that the $SF$s of the first two $0^+$ states
   are summed up to 2.0. The normalized $s$-wave $SF$s for the first two 0$^+$ states are 0.60 $\pm$ 0.13 and 1.40 $\pm$ 0.31. This result indicates that, similar to $^{12}$Be,  more $s$-wave component appears in the excited 0$_2^+$ state (70$\%$) than in the 0$_1^+$ ground state (30$\%$). The $s$-wave strength also
indicates that each state has a large number of $(1s_{1/2})^2$
component, revealing strong mixture between $2s_{1/2}$ and $1d_{5/2}$ orbital
configurations in the low-lying states of $^{16}$C. Noting that the \SI{30}{\percent} $s$-wave component in the ground-state wave function of ${\rm ^{16}{C}}$ is much smaller than that in ${\rm ^{15}{C}}$, but much larger than that in ${\rm ^{17}{C}}$ predominated by $d$-wave. Further experimental studies, such as $d$($^{16}$C, $^3$He)$^{15}$B, $d$($^{17}$N, $^3$He)$^{16}$C  and $d$($^{16}$C, $p$)$^{17}$C \cite{bib:51}, are required to explain the differences between those neutron-rich carbon isotopes .

\subsection{Summary of experimental results}
\begin{figure}[!htb]
\includegraphics
  [width=0.95\hsize]
  {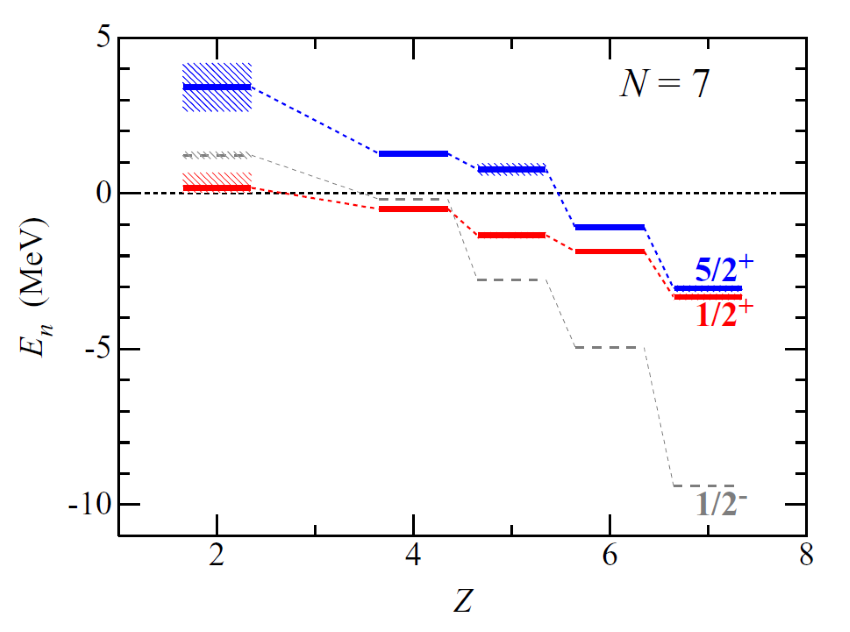}
\caption{The experimental data available on the energy $E_n$, relative to the neutron threshold of $1p_{1/2}$,$1d_{5/2}$ and $2s_{1/2}$ orbitals for $N$ = 7 nuclei.  This figure is from Ref. \cite{bibxx0}.}
\label{fig:n7-fig}
\end{figure}

\begin{figure}[!htb]
\includegraphics
  [width=0.95\hsize]
  {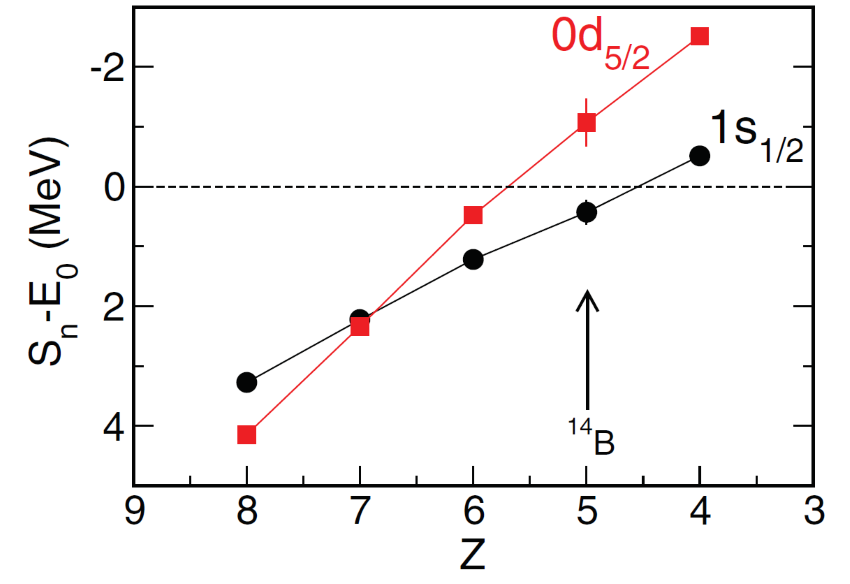}
\caption{Effective single-particle binding energies for $N$ = 9 isotones. This figure is from Ref. \cite{Bedoor}.}
\label{fig:n9-fig}
\end{figure}

Now, a brief summary of experimental studies on the intruder components in neutron-excess He, Li, Be, B, C isotopes will be shown in this subsection.
For $N$ = 7 and $N$ = 9 systems, the variation in energy of $2s_{1/2}$ orbital relative to $1d_{5/2}$ orbital is presented in Fig.~\ref{fig:n7-fig} and Fig.~\ref{fig:n9-fig}, respectively. For the $N$ = 7 isotones, the ordering of $s$ and $d$ orbitals are reversed, and the energy gap between these two orbitals becomes smaller as $Z$ increases. The $s$ orbital also moves rapidly with respect to the $p$ orbital for $N$ = 7 nuclei. The sequences of these two orbitals are inverted in $^{9}$He and $^{11}$Be but are reverted to the normal order in $^{12}$B and $^{13}$C. For $N$ = 9 isotones, the effective single-particle energies for $2s_{1/2}$ and $1d_{5/2}$ orbitals relative to the one-neutron separation energy, clearly shows the inversion of $s$ and $d$ orbital in $^{13}$Be, $^{14}$B and $^{15}$C, and the recovery in $^{16}$N and $^{17}$O. There are two valence neutrons in $N$ = 6, $N$ = 8  and $N$ = 9 nuclei, so the experimental studies of these isotopes are more complicated and the results are not so conclusive except for $^8$He, $^{11}$Li, $^{12}$Be and $^{16}$C. The conclusive results are as follows. In the ground state of $^8$He, the intruder $(1p_{1/2})^2$ component can not be neglected. In the ground state of $^{11}$Li, the intruder $(2s_{1/2})^2$ configurations is as important as the normal $(1p_{1/2})^2$ component. For $^{12}$Be and $^{16}$C,  the $(2s_{1/2})^2$ intrusion appears more in the excited $0_2^+$ state than that in the $0_1^+$ ground state. The core $^{6}$He ($^{10}$He) was found to be partially excited to the 2$^+$ state in the ground state of $^{8}$He ($^{11}$Li).

From these transfer reactions, some intruder configurations have also been found in the low-lying states of light neutron-rich nuclei. An exotic $s$-wave proton intruder state at $E_x$ = 4.83 MeV was observed in $^{13}$B using a proton transfer reaction of $^{12}$Be, and was re-examined by a new $^{14}$C($d$, $^3$He) experiment. The positive-parity low-lying states in $^{13}$B, which are expected to be mixed with the $2s_{1/2}$ and $1d_{5/2}$ orbitals, were studied by several $d$($^{12}$B, $p$) experiments. Some strengths of $s$- and $d$-wave were not observed experimentally in comparison with the shell model predictions. The first excited state at $E_x$ = 0.74 MeV in $^{14}$B was most likely a $s$-wave halo state. The second $3/2^-$ state in $^{15}$B, similar to the $0_2^+$ state in $^{12}$Be and $^{16}$C,  was predicted to be a $s$-wave predominant state, but was not found in experiment. Two single-nucleon transfer reaction experiments, $d$($^{14}$B, $p$) and $d$($^{16}$C, $^3$He), have been performed by nuclear physics experimental group at Peking University in order to search for such an exotic state and study the intruder component in low-lying states of $^{15}$B. The data are under analysis, but there is no conclusion until now.
Summed up, these results show clearly that the experimental studies on intruder component in light neutron-rich nuclei offer an opportunity to understand the shell evolution in exotic nuclei.

Besides, transfer reaction in inverse kinematics has also been used to study the shell evolution in nuclei around the conventional magic number $N$ = 20, 28, 50, 82 as well as the new magic number, such as $N$ = 14, 16 and 32.   Table.~\ref{tab:largern-tab} summarizes the recently reported transfer reaction experiments focusing on heavier nuclei. The experimental results have been partly reviewed in Ref. \cite{Wimmer}. In addition to shell evolution, these experiments were also used to study the proton–neutron asymmetry which is related to short-range correlations \cite{pang-plb}, and to provide the important reaction ratio for the nuclear astrophysics. A summary for the last topic can be found in Ref. \cite{Bardayan}.

\begin{table}[!htb]
\caption{A brief summary of the recently reported transfer reaction experiments focusing on heavier nuclei.}
\label{tab:largern-tab}
\begin{tabular*}{8cm} {@{\extracolsep{\fill} } lllr}
\toprule
Experiment & Facility & Year & Reference \\
\midrule
$d$(${\rm ^{15}{N}}$, $p$)${\rm ^{16}{N}}$ & ORNL & 2008 & \cite{bibn01} \\
$d$(${\rm ^{14}{O}}$, $t$)${\rm ^{13}{O}}$ & GANIL & 2013 & \cite{bib:7} \\
$d$(${\rm ^{14}{O}}$, ${\rm ^{3}{He}}$)${\rm ^{13}{N}}$ & GANIL & 2013 & \cite{bib:7} \\
$d$(${\rm ^{16}{O}}$, $p$)${\rm ^{17}{O}}$ & CIAE & 2019 & \cite{bibn02} \\
$d$(${\rm ^{19}{O}}$, $p$)${\rm ^{20}{O}}$ & ANL & 2012 & \cite{Hoffman-19O} \\
$d$(${\rm ^{20}{O}}$, $p$)${\rm ^{21}{O}}$ & GANIL & 2011 & \cite{bib:8} \\
$d$(${\rm ^{22}{O}}$, $p$)${\rm ^{23}{O}}^{*}$ & GANIL & 2007 & \cite{bib:61} \\
$d$($^{18}{\rm {F}}$, $p$)${\rm ^{19}{F}}$ & ANL & 2018 & \cite{bib:25} \\
$d$(${\rm ^{21}{F}}$, $p$)${\rm ^{22}{F}}$ & ANL & 2018 & \cite{bibn03} \\
$d$(${\rm ^{19}{Ne}}$, $n$)${\rm ^{20}{Na}}$ & FSU & 2016 & \cite{bib:31} \\
$d$(${\rm ^{24}{Ne}}$, $p$)${\rm ^{25}{Ne}}$ & GANIL & 2010 & \cite{bib:12} \\
$d$(${\rm ^{26}{Ne}}$, $p$)${\rm ^{27}{Ne}}$ & GANIL & 2012 & \cite{bibn04} \\
$d$(${\rm ^{25}{Na}}$, $p$)${\rm ^{26}{Na}}$ & TRIUMF & 2016 & \cite{bibn05} \\
$d$(${\rm ^{28}{Mg}}$, $p\gamma$)${\rm ^{29}{Mg}}$ & TRIUMF & 2019 & \cite{bibn06} \\
$d$(${\rm ^{26}{Al}}^{m}$, $p$)${\rm ^{27}{Al}}$ & ANL & 2017 & \cite{bib:26} \\
$d$(${\rm ^{34}{Si}}$, $p$)${\rm ^{35}{Si}}$ & GANIL & 2014 & \cite{bib:10} \\
$p$(${\rm ^{34,46}{Ar}}$, $d$) & NSCL & 2010 & \cite{bib:38} \\
$d$(${\rm ^{44}{Ar}}$, $p$)${\rm ^{45}{Ar}}$ & GANIL & 2008 & \cite{bibn07} \\
$d$(${\rm ^{46}{Ar}}$, $p$)${\rm ^{47}{Ar}}$ & GANIL & 2006 & \cite{bibn08} \\
$d$(${\rm ^{60}{Fe}}$, $p$)${\rm ^{61}{Fe}^{*}}$ & GNAIL & 2017 & \cite{bib:9} \\
$p$(${\rm ^{56}{Ni}}$, $d$)${\rm ^{55}{Ni}}$ & NSCL & 2014 & \cite{bib:39} \\
$d$(${\rm ^{66}{Ni}}$, $p$)${\rm ^{67}{Ni}}$ & CERN & 2014 & \cite{bib:19} \\
$d$(${\rm ^{66}{Ni}}$, $p$)${\rm ^{67}{Ni}}$ & CERN & 2015 & \cite{bibn09} \\
$d$(${\rm ^{70}{Zn}}$, ${\rm ^{3}{He}}$)${\rm ^{69}{Cu}}$ & Alto & 2016 & \cite{bibn10} \\
$d$(${\rm ^{72}{Zn}}$, ${\rm ^{3}{He}}$)${\rm ^{71}{Cu}}$ & GANIL & 2015 & \cite{bibn11} \\
$d$(${\rm ^{78}{Zn}}$, $p$)${\rm ^{79}{Zn}}$ & CERN & 2015 & \cite{bib:18} \\
$d$(${\rm ^{82}{Ge}}$, $p$)${\rm ^{83}{Ge}}$ & ORNL & 2005 & \cite{bibn12} \\
$d$(${\rm ^{84}{Se}}$, $p$)${\rm ^{85}{Se}}$ & ORNL & 2005 & \cite{bibn12} \\
$d$(${\rm ^{95}{Sr}}$, $p$)${\rm ^{96}{Sr}}$ & TRIUMF & 2018 & \cite{bibn13} \\
$d$(${\rm ^{94,95,96}{Sr}}$, $p$) & TRIUMF & 2019 & \cite{bibn14} \\
$d$(${\rm ^{132}{Sn}}$, $p$)${\rm ^{133}{Sn}}$ & ORNL & 2010 & \cite{bibn15} \\
$d$(${\rm ^{132}{Sn}}$, $t$)${\rm ^{131}{Sn}}$ & ORNL & 2018 & \cite{bibn16} \\
$p$(${\rm ^{154,159}{Gd}}$, $d$) & LBNL & 2014 & \cite{bibn17} \\
\bottomrule
\end{tabular*}
\end{table}

\section{Summary} \label{summary}

Single-nucleon transfer reactions in inverse kinematics is a sensitive tool to investigate the exotic structure in unstable neutron-rich nuclei. The DCSs (or angular distributions) of the selectively populated final states contain essential structure information to help us understand the nature of nuclear force and interactions in unstable nuclei. The transferred momentum $l$ (or spin-parity of low-lying states in final nucleus), the $SF$s, and the effective single-particle energy can be deduced from the single-particle transfer reaction. These observables are very useful to interpret the shell evolution in neutron-excess nuclei. The inverse kinematics rather than the normal kinematics is preferred to be used in transfer reactions with radioactive beams. Therefore, many experimental setups including silicon detection arrays (with or without $\gamma$-ray array), a special spectrometer HELIOS, as well as newly developed AT-TPC, which are appropriate to be used in single-nucleon transfer reaction in inverse kinematics, were developed at various laboratories around the world. The basic modules, advantages and disadvantages of each array are illustrated.

A large number of single-nucleon transfer reaction experiments were performed with radioactive beams in order to quantitatively study the intruder components in low-lying states of neutron-rich He, Li, Be, B and C isotopes. For the ground state of the $N$ = 7 and $N$ = 9 isotones, the experimental results are conclusive. For example, the intruder $s$-wave component dominates in $^9$He, $^{10}$Li (from other reactions), $^{11}$Be,  $^{14}$B and $^{15}$C. However, for the ground state of the $N=6$, $N=8$ and $N=10$ isotones, the exsistence of two valence neutrons leads to more complications. Thus, some experimental results are disputed, but one still can conclude something on the ground state of $^{8}$He, $^{11}$Li, $^{12}$Be and $^{16}$C. First of all, the intruder $(1p_{1/2})^2$ component in $^{8}$He can not be ignored.  Moreover, the strengths of $(2s_{1/2})^2$ intrusion in $^{11}$Li, $^{12}$Be, and $^{16}$C are 47$\%$, 19$\%$ and 30$\%$, respectively, while the intensities of $(1p_{1/2})^2$ in $^{11}$Li as well as $(1d_{5/2})^2$ in $^{12}$Be and $^{16}$C are larger than 50$\%$. The $s$-wave intrusions were also found in some excited states, such as the proton intruder state in $^{13}$B, the first excited state in $^{14}$B, the second 0$^+$ states in $^{12}$Be and $^{16}$C.
All these experimental studies and conclusions are helpful in explaining the anomalous behaviour of the conventional magic number $N$ = 8 in light neutron-rich nuclei. With the development of accelerators and detection technique, it can be  anticipated that more single-particle transfer reactions will be performed in the future.

\end{document}